\documentclass[useAMS,usenatbib]{mn2e}

\usepackage{graphicx,amssymb,amstext} 
\usepackage{hyperref}

\def\aj{AJ}			
		
\def\apj{ApJ}		
\def\apjl{ApJ}		
\def\apjs{ApJS}				
		
\def\aap{A\&A}

\def\mnras{MNRAS}

\def\pasp{PASP}

\begin{document}

 \title[The Revised IRAS-FSC Redshift Catalogue]{The  {\it Revised IRAS-FSC Redshift Catalogue} (RIFSCz)} 
\author[L. Wang et al.]{
\parbox[t]{\textwidth}{
Lingyu Wang$^{1}$\thanks{E-mail: lingyu.wang@durham.ac.uk}, Michael Rowan-Robinson$^{2}$, Peder Norberg$^{1}$, Sebastien Heinis$^{3}$, Jiaxin Han$^{1}$}
\\
\\
$^{1}$Institute for Computational Cosmology, Department of Physics, Durham University, South Road, Durham, DH1 3LE, UK\\
$^{2}$Astrophysics Group, Blackett Laboratory, Imperial College of Science Technology and Medicine, London SW7 2BZ, UK\\
$^{3}$Department of Astronomy, University of Maryland, College Park, MD 20742-2421, USA\\
}

\date{Accepted . Received ; in original form }

\maketitle

\begin{abstract}  
We present a {\it Revised IRAS-FSC Redshift Catalogue} (RIFSCz) of 60,303 galaxies selected at 60\ $\micron$ from the {\it IRAS Faint Source Catalogue} (FSC). This revision merges in data from the WISE All-Sky Data Release, the Tenth SDSS Data Release (DR10), the GALEX  All-Sky Survey Source Catalog (GASC), the 2MASS Redshift Survey (2MRS) and the Planck Catalogue of Compact Sources (PCCS). The RIFSCz consists of accurate position, ultra-violet (UV), optical, near-, mid- and far-infrared, sub-millimetre (sub-mm) and/or radio identifications, spectroscopic redshift (if available) or photometric redshift (if possible), predicted far-infrared and sub-mm fluxes ranging from 12 to 1380\ $\micron$ based upon the best-fit infrared template. We also provide stellar masses, star-formation rates and dust masses derived from the optical and infrared template fits, where possible. $56\%$ of the galaxies in the RIFSCz have spectroscopic redshifts and a further $26\%$ have photometric redshifts obtained through the template-fitting method. At S60 $>$ 0.36 Jy, the 90$\%$ completeness limit of the FSC, $93\%$ of the sources in the RIFSCz have either spectroscopic or photometric redshifts. An interesting subset of the catalogue is the sources detected by Planck at sub-mm wavelengths. 1200 sources have a detection at better than 5-$\sigma$ in at least one Planck band and a further 1186 sources have detections at 3-5$\sigma$ in at least one Planck band. 


\end{abstract}


\section{INTRODUCTION}

The {\it IRAS Faint Source Catalog} (FSC; Moshir et al. 1992) contains 173,044 sources reaching a depth of $\sim$0.2 Jy at 12, 25 and 60\ $\micron$. It is limited to $|b|>20^\circ$ in unconfused regions at 60\ $\micron$. For sources with high-quality flux density\footnote{In the IRAS FSC, the flux density quality (FQUAL) is classified as high (=3), moderate (=2) or upper limit (=1)}, reliability is $>99\%$ at 12 and 25\ $\micron$ and $>94\%$ at 60\ $\micron$. Around 41$\%$ of the FSC sources are detected at 60 $\mu$m (FQUAL $>$ 1). The construction of  the Imperial IRAS-FSC Redshift Catalogue (IIFSCz; Wang \& Rowan-Robinson 2009) was made possible by overlaps (in terms of depth and area) with various imaging and/or redshift surveys, such as the Sloan Digital Sky Survey (SDSS; York et al. 2000), the Two Micron All Sky Survey (2MASS; Skrutskie et al. 2006) and the 6dF Galaxy Survey (Jones et al. 2004, 2005). However, the large positional error of the IRAS sources has meant in particular that the cross-identification between IRAS and the deep optical data SDSS is very challenging. The completion and/or release of several major surveys in the last few years, including the Wide-field Infrared Survey Explorer (WISE; Wright et al. 2010) All-Sky Survey, the GALEX All-Sky Survey Source Catalog (GASC; Seibert et al., in prep.), the 2MASS Redshift Survey (2MRS; Huchra et al. 2012), the Tenth SDSS Data Release (DR10; Ahn et al. 2013) and the Planck all-sky survey (Planck Collaboration I 2013) in the microwave and sub-millimetre (sub-mm), has made it imperative to revise the IIFSCz. In particular, the imaging depth and sky coverage of WISE are such that the WISE all-sky source catalogue provides counterparts for the majority of sources in the IIFSCz and thereby improves the positional accuracy of these sources by a factor of $\sim$10. Using the WISE positions, cross-identification between IRAS sources and sources detected at UV, optical and near-infrared becomes much easier and much more reliable. 

The layout of this paper is as follows. In Section~\ref{sec:identification}, first we cross-identify IRAS FSC sources with their mid-infrared (MIR) WISE counterparts using a likelihood ratio technique. Then, using their WISE positions, we cross-match FSC sources with sources detected in other wavebands including the ultra-violet (UV), optical, near-infrared (NIR) and sub-mm.  Spectroscopic redshifts are collected from a number of databases, e.g., SDSS DR10, 2MRS and NED. In Section 3, we estimate photometric redshift using a template-fitting method for sources matched with WISE, SDSS and/or 2MASS counterparts but do not have spectroscopic redshifts. In Section 4,  infrared templates are fitted to the mid- and far-infrared data from IRAS and WISE as well as sub-mm data from Planck for FSC sources with WISE and Planck associations. Once a best-fit infrared template is found, we make flux predictions at far-infrared and sub-mm wavelengths.  Finally,  discussions and conclusions of the overall properties of the revised IRAS-FSC Redshift  catalogue (RIFSCz) are given in Section 5. Throughout the paper, we adopt a flat cosmological model with $\Lambda=0.7$ and $h_0=0.72$ and a Salpeter initial mass function (Salpeter et al. 1955). Unless otherwise stated, we use the AB magnitude system, and $\log=\log_{10}$.

\section{SOURCE IDENTIFICATION}
\label{sec:identification}

Our starting point is the complete sample of galaxies selected at 60\ $\micron$ from the IRAS FSC, presented in Wang \& Rowan-Robinson (2009). For the sake of completeness, we briefly summarise our selection criteria here: (1) To ensure reliability, we select sources with FQUAL $\geq3$ and SNR $>5$ at 60\ $\micron$;  (2) To exclude cirrus, we require $\log$(S100/S60)$<0.8$ if FQUAL $\geq2$ at 100\ $\micron$; (3) To discriminate against stars, we firstly require $\log$(S60/S25)$>-0.3$ if FQUAL $\geq2$ at 25\ $\micron$ and then $\log$(S60/S12)$>0$ if FQUAL $\geq$ 2 at 12\ $\micron$.  

\subsection{Cross-match with WISE}

 
The Wide-field Infrared Survey Explorer (WISE; Wright et al. 2010) mapped the sky at 3.4, 4.6, 12, and 22\ $\micron$ (W1, W2, W3, W4) with an angular resolution of 6.1\arcsec, 6.4\arcsec, 6.5\arcsec and 12.0\arcsec respectively. The All-Sky Release includes all data taken during the full cryogenic mission phase and the All-Sky Source Catalog contains the properties of over 563 million point-like and resolved objects.  To ensure a high degree of reliability, sources are required to meet SNR $>$ 5 in at least one WISE band and other criteria.   Photometry is presented in the form of point source profile-fitting measurement and multi-aperture photometry.  WISE 5$\sigma$ photometric sensitivity is estimated to be 0.068, 0.098, 0.86 and 5.4 mJy at 3.4, 4.6, 12 and 22\ $\micron$ in unconfused regions on the ecliptic plane (Wright et al. 2010). Sensitivity is better in regions at higher ecliptic latitudes with deeper coverage and lower zodiacal background, and worse in regions with high source density or complex background. WISE provides Vega magnitudes and we have taken the corrections to AB magnitudes to be 2.683, 3.319, 5.242 and 6.604 mags at 3.4, 4.6, 12 and 22\ $\micron$.

\subsubsection{The likelihood ratio technique}


\begin{figure}
\includegraphics[height=2.57in,width=3.4in]{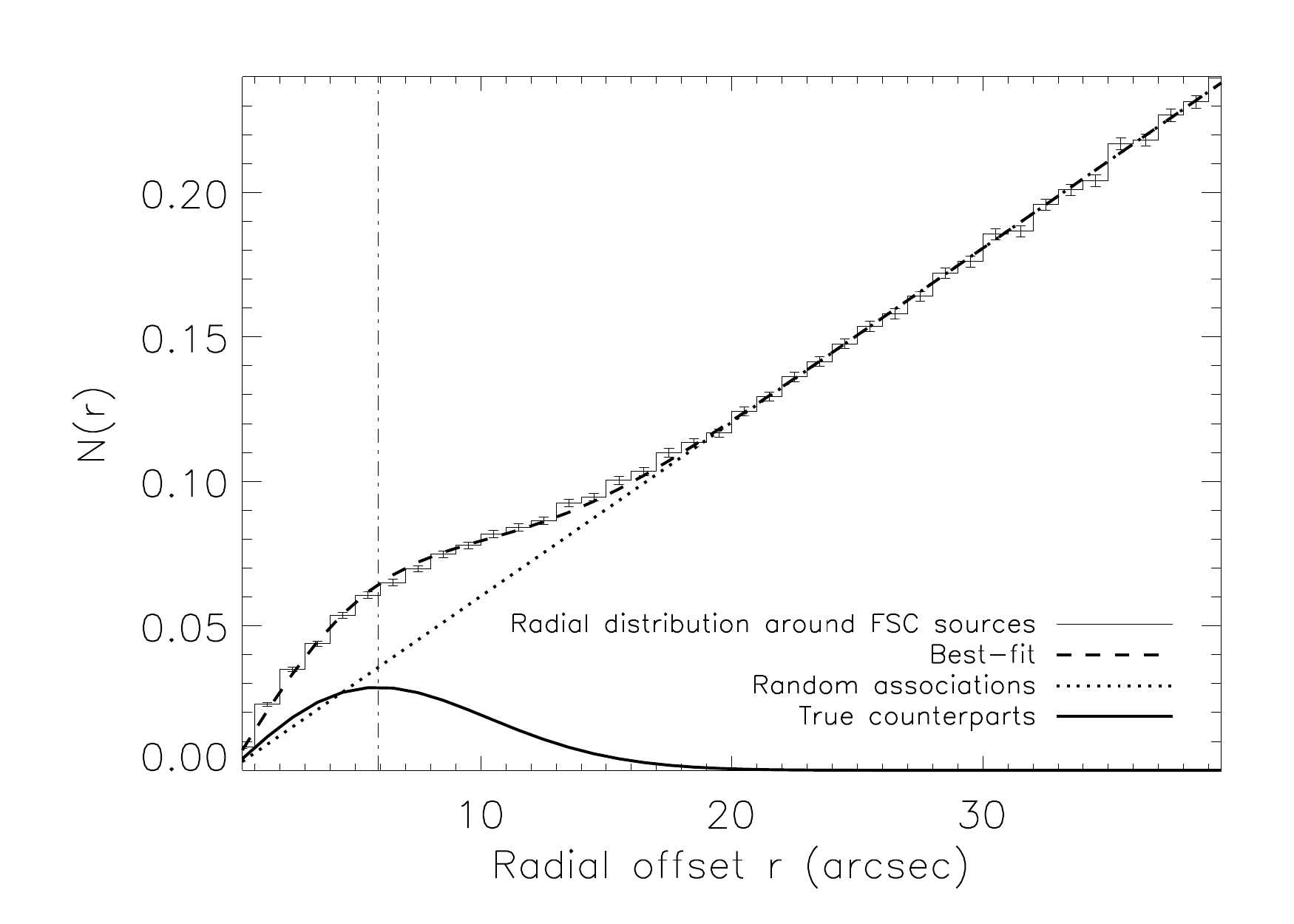}
\includegraphics[height=2.57in,width=3.4in]{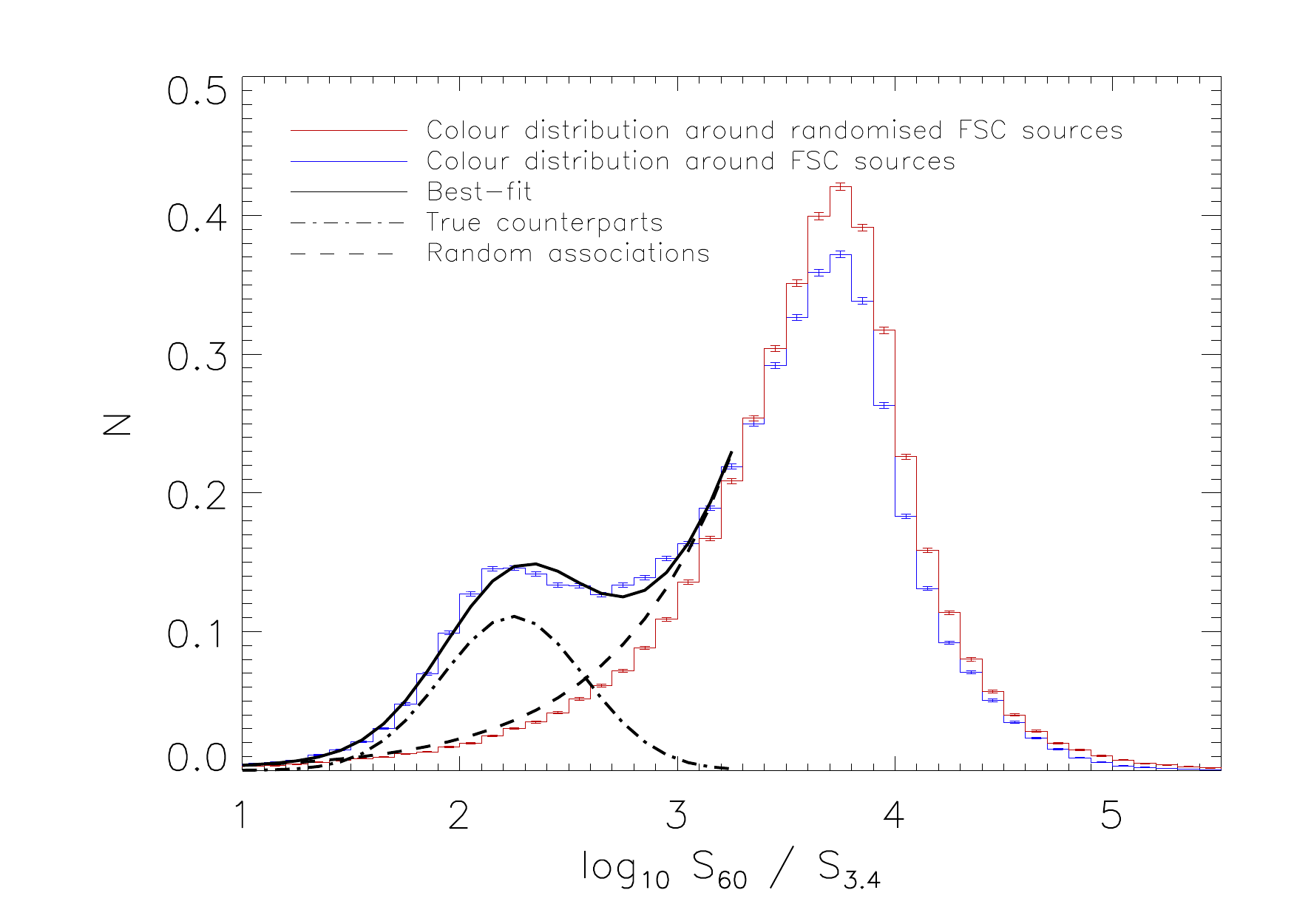}
\caption{Top: The average distribution of radial offsets between FSC and WISE sources per FSC source (black histogram) which contains both true FSC - WISE matches and random WISE sources uncorrelated with FSC sources. The radial distribution of the random associations follows a linear trend with $r$ (dotted line), while the radial distribution of the true counterparts follows the Rayleigh radial distribution (solid line). The dashed line corresponds to the sum of the solid line and the dotted line. The vertical line marks $\sigma_r=5.78\arcsec$ where the radial distribution of the true counterparts peaks. Bottom: The average distribution of the 60-to-3.4\ $\micron$ colour of all WISE sources within 40\arcsec\ per FSC source (blue histogram) compared to that of all WISE sources within 40\arcsec\ of a random location (red histogram). We assume that the colour distribution of the true counterparts can be fit by a Gaussian distribution (dot-dashed line) and the colour distribution of the random associations can be fit by an exponential function (dashed line). The solid line is the sum of the dashed line and the dot-dashed line. To avoid incompleteness problems at faint 3.4\ $\micron$ fluxes (i.e. large 60-to-3.4\ $\micron$ colour), the fitting is carried out at $\log_{10}(S_{60}/S_{3.4})<3.4$.}
\label{fig:offset}
\end{figure}

\begin{figure}
\includegraphics[height=2.5in,width=3.45in]{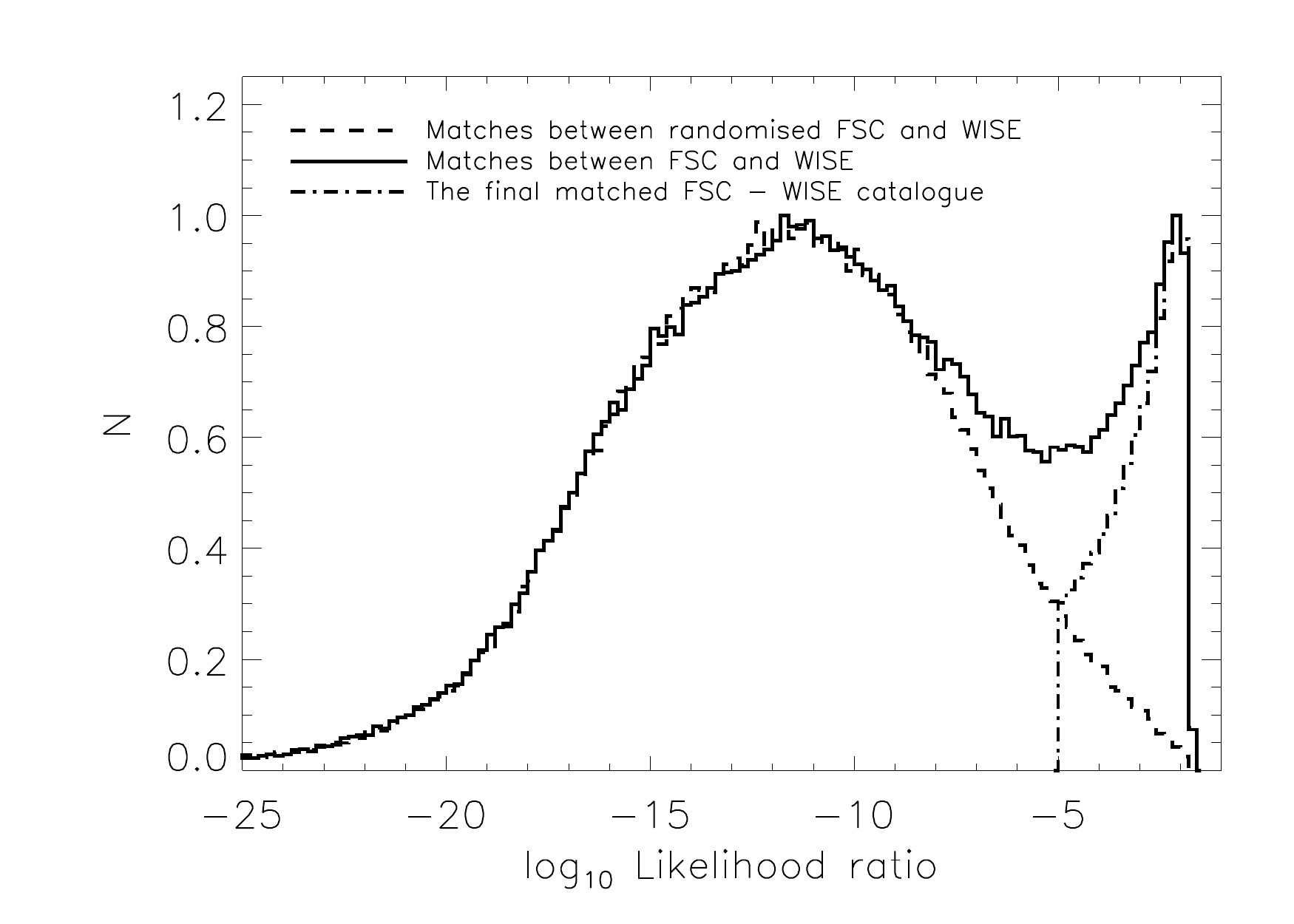}
\caption{The normalised LR distribution for all  matches between FSC andWISE (solid line), all matches between the randomised FSC and WISE (dashed line) and the final cross-matched FSC-WISE catalogue (dot-dashed line)e. The exact value of the likelihood ratio is not important, only the relative ordering matters.}
\label{fig:fid_comp}
\end{figure}

\begin{figure*}
\includegraphics[height=2.123in,width=2.3009in]{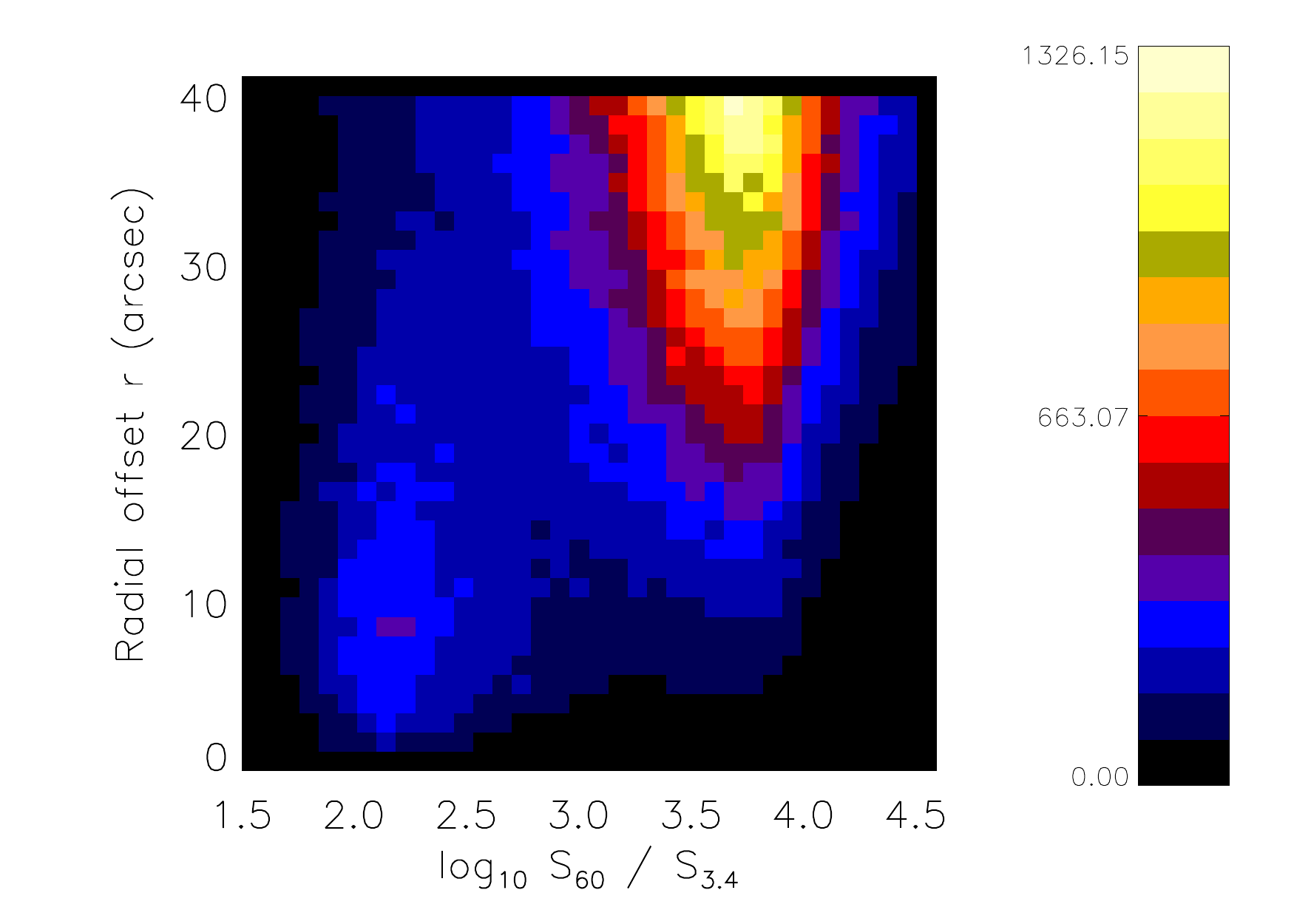}
\includegraphics[height=2.123in,width=2.3009in]{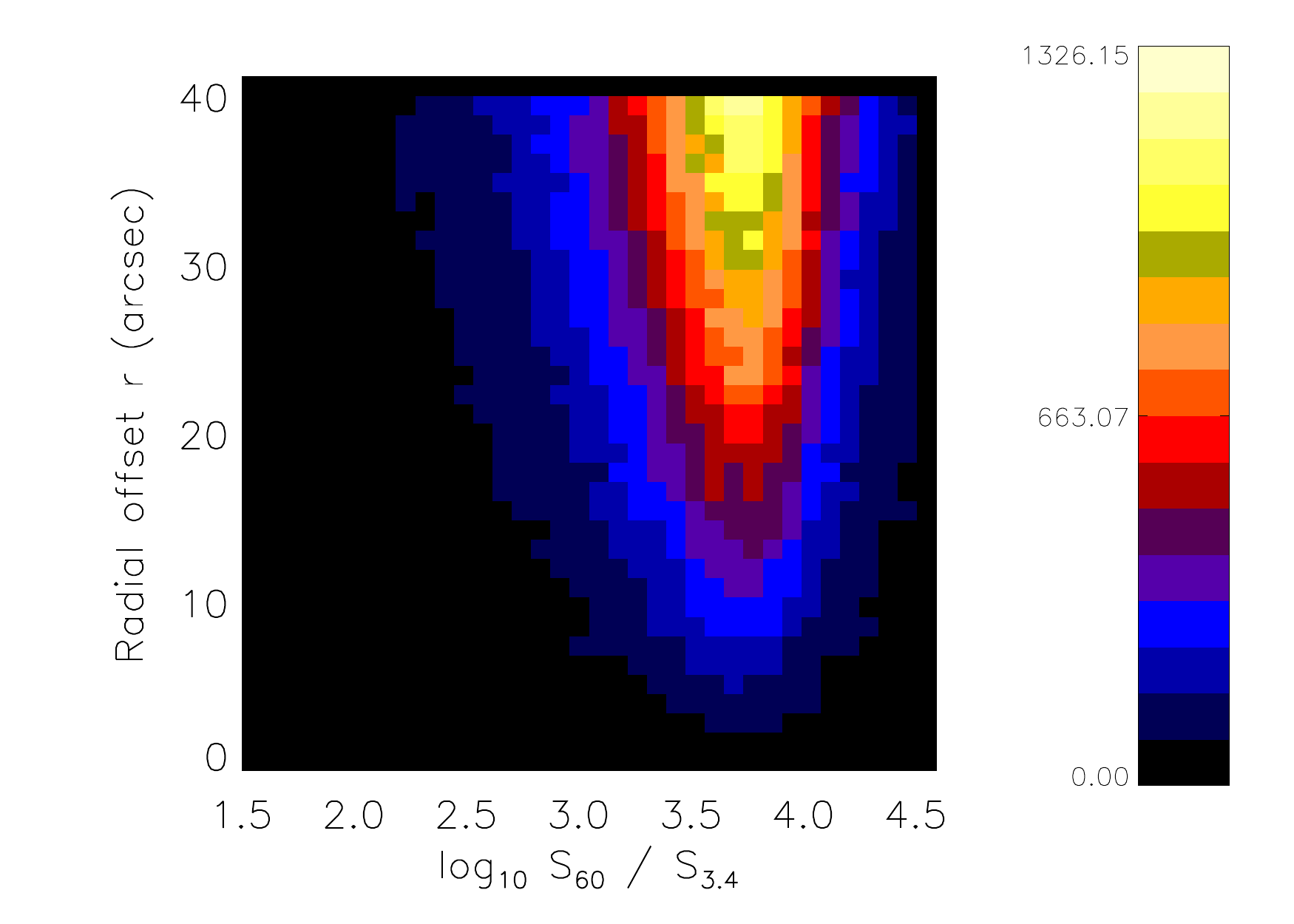}
\includegraphics[height=2.123in,width=2.3009in]{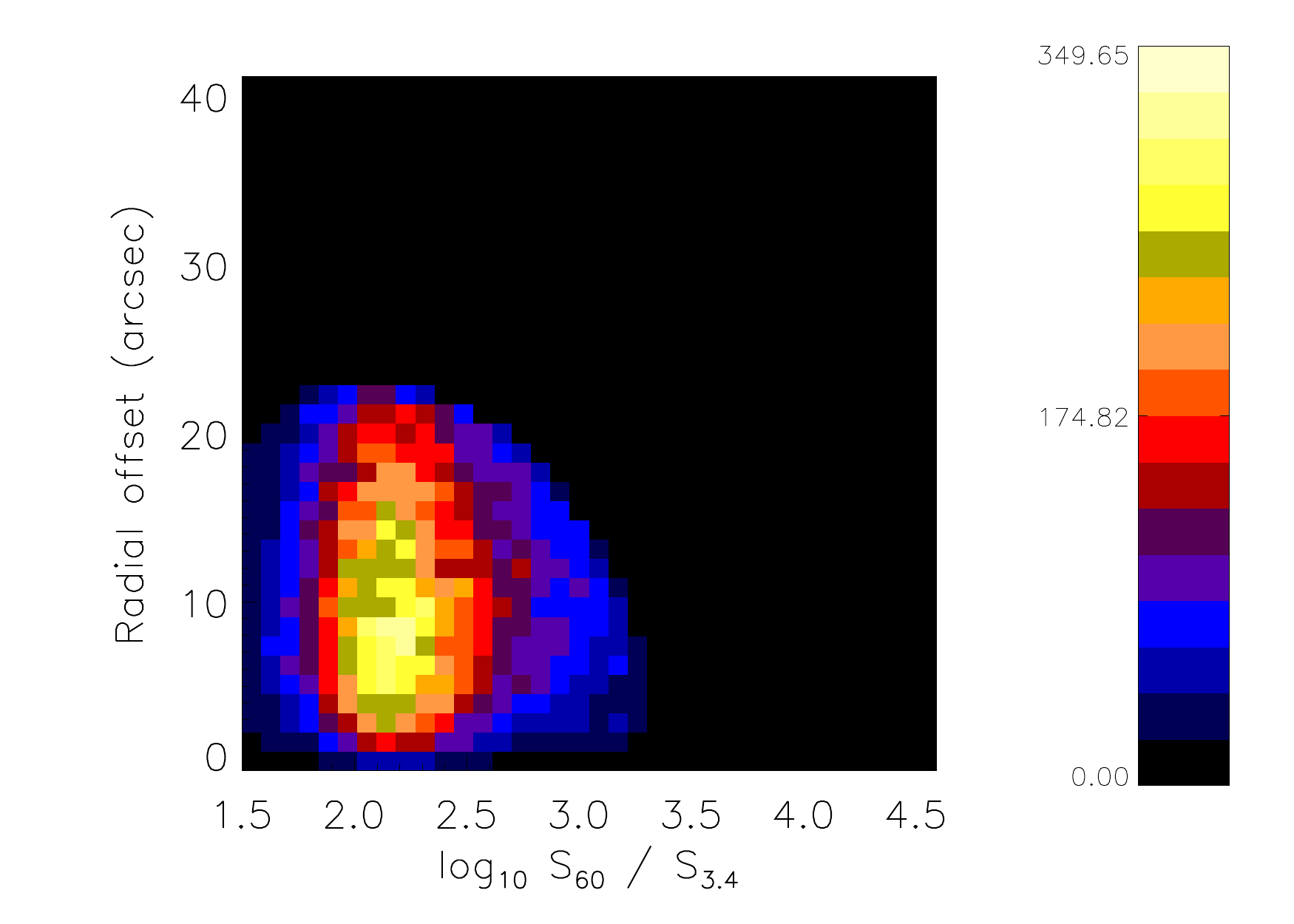}
\caption{Left:  The 2D distribution in $S_{60}/S_{3.4}$ and radial offset $r$ of all matches between FSC and WISE, which contains both real FSC-WISE matches and random associations. The colour coding is based on the number of matches in a given cell in the positional offset versus colour plane. Middle: The 2D distribution of all matches between randomised FSC and WISE, which contains only random associations. Right:  The 2D distribution of the final matched FSC-WISE catalogue of 48,603 sources, using the likelihood ratio technique.}
\label{fig:2dratio}
\end{figure*}


We use the likelihood ratio technique (LR; Sutherland \& Saunders 1992; Brusa et al. 2007;  Wang \& Rowan-Robinson 2010; Chapin et al. 2011; Wang et al. 2013) to cross-match FSC sources with counterparts at 3.4\ $\micron$ which is the most sensitive WISE band. The LR technique compares the probability of a true counterpart with the probability of a chance association, as a function of 60-to-3.4\ $\micron$ flux ratio $S_{60} / S_{3.4}$ and radial offset $r$ . Assuming the probability of true counterpart and random association is separable in $\log_{10}(S_{60} / S_{3.4})$ (or $C_{60-3.4}$ as a shorthand) and $r$, we can write
\begin{equation}
LR=\frac{{\rm Prob}_{\rm true}(C_{60-3.4}, r)}{{\rm Prob}_{\rm random}(C_{60-3.4}, r)}=\frac{q(C_{60-3.4}) f( r )dCdr}{p(C_{60-3.4}) b( r ) dCdr},
\end{equation}
where $q(C_{60-3.4})$ and $p(C_{60-3.4})$ are the colour distributions of the true counterparts and random matches respectively, and $f( r )$ and $b( r )$ are the positional distributions of the true counterparts and random associations respectively.
To derive the positional distribution of the true counterparts $f( r )$, we assume a symmetric Gaussian distribution as a function of orthogonal positional coordinates. Therefore,  $f ( r )$ can be written as  a Rayleigh radial distribution,
\begin{equation}
f( r ) dr = \frac{r}{\sigma_r^2} \exp(-r^2 / 2 \sigma_r^2) dr,
\end{equation}
where the scale parameter of the Rayleigh distribution $\sigma_r$ is where $f( r )$ peaks and $\int_0^\infty f( r ) dr = 1$.
The positional distribution of random associations can be written as 
\begin{equation}
b( r ) dr = 2\pi r dr,
\end{equation}
assuming a constant surface density of random background 3.4\ $\micron$ sources uncorrelated with FSC sources. In the top panel in Fig.~\ref{fig:offset}, we plot the average distribution of radial offsets between FSC and WISE sources per FSC source, which contains both the true counterparts and the random associations. We fit our model 
\begin{equation}
N( r ) dr = E \times f( r )  dr + \rho \times b ( r ) dr
\end{equation}
to the observed histogram to determine the best-fit parameters to be $E=0.30\pm0.01$, $\sigma_r=5.78\arcsec\pm0.20\arcsec$, and $\rho=0.93\pm0.03$ arc sec$^{-2}$. Consequently, the best-fit radial distribution of the true counterparts and that of the random associations can be characterised using our model, which are plotted as the thick solid line and the dotted line in the top panel in Fig.~\ref{fig:offset}. In the bottom panel in Fig.~\ref{fig:offset}, we plot the average 60-to-3.4\ $\micron$ colour distribution of all FSC-WISE matches per FSC source and per randomised FSC source. The former contains both real FSC-WISE matches and random associations. The latter contains only random 3.4\ $\micron$ sources uncorrelated with FSC sources. We generate the randomised FSC catalogue by locally randomising the IRAS positions and randomly swapping the 60\ $\micron$ flux densities. The colour distribution per randomised FSC source can be characterised by an exponential increase towards larger values of $\log_{10}(S_{60}/S_{3.4})$ (fainter 3.4\ $\micron$ flux densities) followed by an exponential decrease due to incompleteness. The colour distribution per FSC source can be characterised by a Gaussian distribution of the true counterparts and the double exponential form of the random associations. Consequently, we fit a Gaussian and an exponential function to the colour distribution per FSC source at $\log_{10}(S_{60}/S_{3.4})<3.4$.

We can now calculate the likelihood ratio for every possible match between FSC and WISE sources.  The exact value of the likelihood ratio is not important, only the relative ordering matters. In Fig.~\ref{fig:fid_comp}, we plot the LR distribution of all matches between FSC and WISE (the solid histogram) and the LR distribution of all matches between the randomised FSC and WISE (the dashed histogram). At $\log_{10} LR < -9$, the solid histogram becomes indistinguishable from the dashed histogram, which indicates that the vast majority of the real matches between FSC and WISE have $\log_{10} LR < -9$.  It is clear that real matches between FSC and WISE sources form a peak at high LR values.  We select FSC-WISE matches with $\log_{10} {\rm LR}\ge -5$ which corresponds to $\sim$4\% false identification rate\footnote{The false identification rate is the ratio of the number of matches between the randomised FSC and WISE above the chosen LR threshold to the total number of matches.}. In cases where there are more than one candidate WISE source matched to the same FSC source\footnote{For all FSC sources matched with WISE counterparts with $\log_{10} {\rm LR}\ge -5$, $25\%$ of them have more than one WISE match. However, we emphasise that this percentage depends on the chosen LR threshold.}, the one with the highest likelihood ratio is selected.  In total, 48,603 IRAS sources ($81\%$ of the RIFSCz catalogue) are matched with a WISE 3.4\ $\micron$ counterpart.  In Fig.~\ref{fig:2dratio}, we plot the 2D distribution in $S_{60}/S_{3.4}$ and radial offset $r$ of all matches between FSC and WISE (left panel), all matches between randomised FSC and WISE (middle), and the final matched FSC-WISE catalogue using the likelihood ratio technique (right).

\begin{figure}
\includegraphics[height=2.6in,width=3.45in]{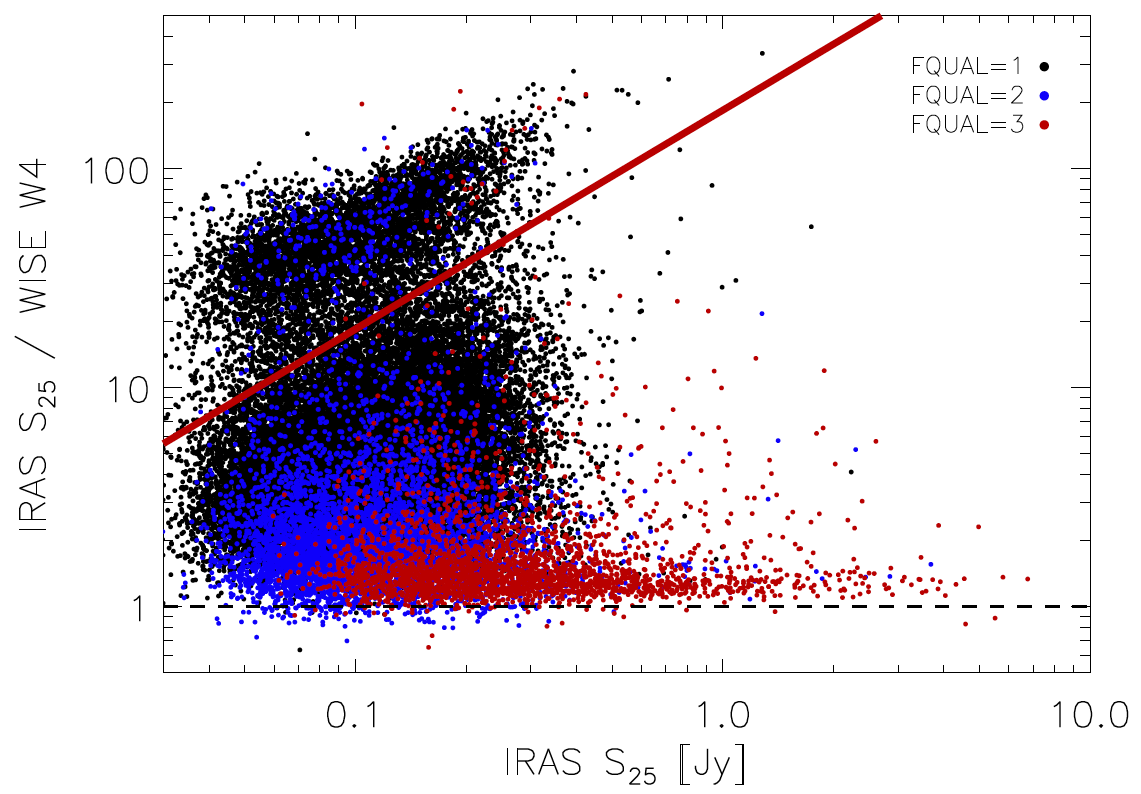}
\caption{IRAS 25\ $\micron$ flux density versus WISE 22\ $\micron$ flux density for FSC - WISE matched sources. A good agreement between WISE and IRAS can be seen for sources with high quality IRAS flux density measurements (FQUAL=3). Above the solid red line, sources are below the WISE 5$\sigma$ photometric sensitivity at 22\ $\micron$.}
\label{fig:iras_vs_wise}
\end{figure}

We can also check the degree of agreement between IRAS 25\ $\micron$ and WISE 22\ $\micron$ detections. In Fig.~\ref{fig:iras_vs_wise}, we plot the IRAS 25\ $\micron$ flux density to WISE 22\ $\micron$ flux density ratio versus the IRAS 25\ $\micron$ flux density for FSC-WISE matched sources with SNR $>$ 2 in the WISE 22\ $\micron$ band and FQUAL in the IRAS 25\ $\micron$ band = 1, 2 or 3.  A good agreement between WISE and IRAS can be seen for sources with high quality IRAS flux density measurements (FQUAL $=3$). WISE fluxes are systematically below the IRAS fluxes even after taking into account the band difference. This systematic trend can be removed after making aperture correction at the WISE bands (see Section 3.1). A similar comparison between the IRAS 12\ $\micron$ fluxes and the WISE 12\ $\micron$ fluxes can be made. However, the number of IRAS sources with high quality flux measurement at 12\ $\micron$ is very small. Therefore, we do not show a plot of the 12\ $\micron$ flux comparison here.



\subsection{Cross-match with GALEX}

GALEX has undertaken a number of surveys covering large areas of sky at a variety of depths. The largest two are the All-Sky (AIS) and Medium Imaging Surveys (MIS). The AIS is much shallower than the MIS but it covers a much larger area of sky. The standard GALEX database contains all of the detected sources which include many duplicate observations of the same source as well as numerous spurious sources. To address these problems, two catalogues of GALEX measurements, namely the GALEX All-Sky Survey Source Catalog (GASC) and the GALEX Medium Imaging Survey Catalog (GMSC), have been constructed (Seibert et al.  in prep.). Sources are selected if they are detected in the NUV with SNR $>3$. Covering a total of 26,300 deg$^2$ of sky, the GASC consists of all GALEX observations reaching a depth of NUV 21 (AB mag) and a total of 40 million unique sources. Following Budav{\'a}ri et al. (2009) and Heinis et al. (2009), we match GALEX sources with FSC sources which have received accurate positions from WISE, using a search radius of 5\arcsec. A total of 31,552 matches are found which correspond to 31,537 unique FSC sources. Only about 10 FSC sources have multiple GALEX matches. For these 10 sources, we simply select the closest GALEX source as the correct association. 

\subsection{Cross-match with SDSS DR10}

The original IIFSCz cross-matched IRAS FSC with SDSS DR6. The source identification process was complicated as the probability of chance association is $\propto n \pi d^2$, where n is the number density of background optical objects and $d$ is the cross-matching radius. The positional uncertainty  of the FSC sources matched with WISE improved drastically thanks to the much higher WISE angular resolution. Following Yan et al. (2013), we carried out source matching between IRAS sources with WISE positions and SDSS DR10 using a matching radius of 3\arcsec.  The SDSS photometric DR10 survey has covered a total unique area of 14,555 deg$^2$ (Ahn et al. 2013).  In total, 25,553 SDSS sources are found within 3\arcsec\ of the IRAS sources matched with WISE. Only primary SDSS objects\footnote{Whenever the SDSS makes multiple observations of the same object, the one with the best photometry will be assigned as the `primary' observation.} are selected.  20,277 FSC sources have unique SDSS association within 3\arcsec , 2,531 FSC sources have multiple SDSS matches. We simply assign the closest SDSS object as the correct optical association for FSC sources with multiple SDSS matches. For some large, nearby galaxies this procedure can lead to association with the wrong SDSS source.  We have corrected this where possible.




\subsection{Cross-match with 2MASS}


The 2MASS survey has imaged the whole sky in the J (1.24\ $\micron$), H (1.66\ $\micron$) and K$_{\rm s}$ (2.16\ $\micron$) near-infrared wavebands (Skrutskie et al. 2006). The 2MASS All-Sky Release Point Source Catalog (PSC) contains attributes in the three survey bands for over 470 million sources. The 2MASS All-Sky Data Release Extended Source Catalog (XSC; Jarrett et al. 2000) contains properties for over 1.6 million objects which are resolved relative to a single point-spread-function.  The PSC contains point-source processed photometry for almost all of the resolved sources in the XSC.

Association information cross-referencing WISE sources with the 2MASS PSC is provided in the WISE All-Sky Source Catalog. In total, there are 45,892 sources in the 2MASS PSC matched to the 48,603 FSC sources which have received WISE positions. We also matched the 48,603 FSC sources with the 2MASS XSC. In total, 35,772 out of the 48,603 FSC sources found a 2MASS XSC counterpart.

\subsection{Cross-match with Planck}

\begin{figure}
\includegraphics[height=2.5in,width=3.4in]{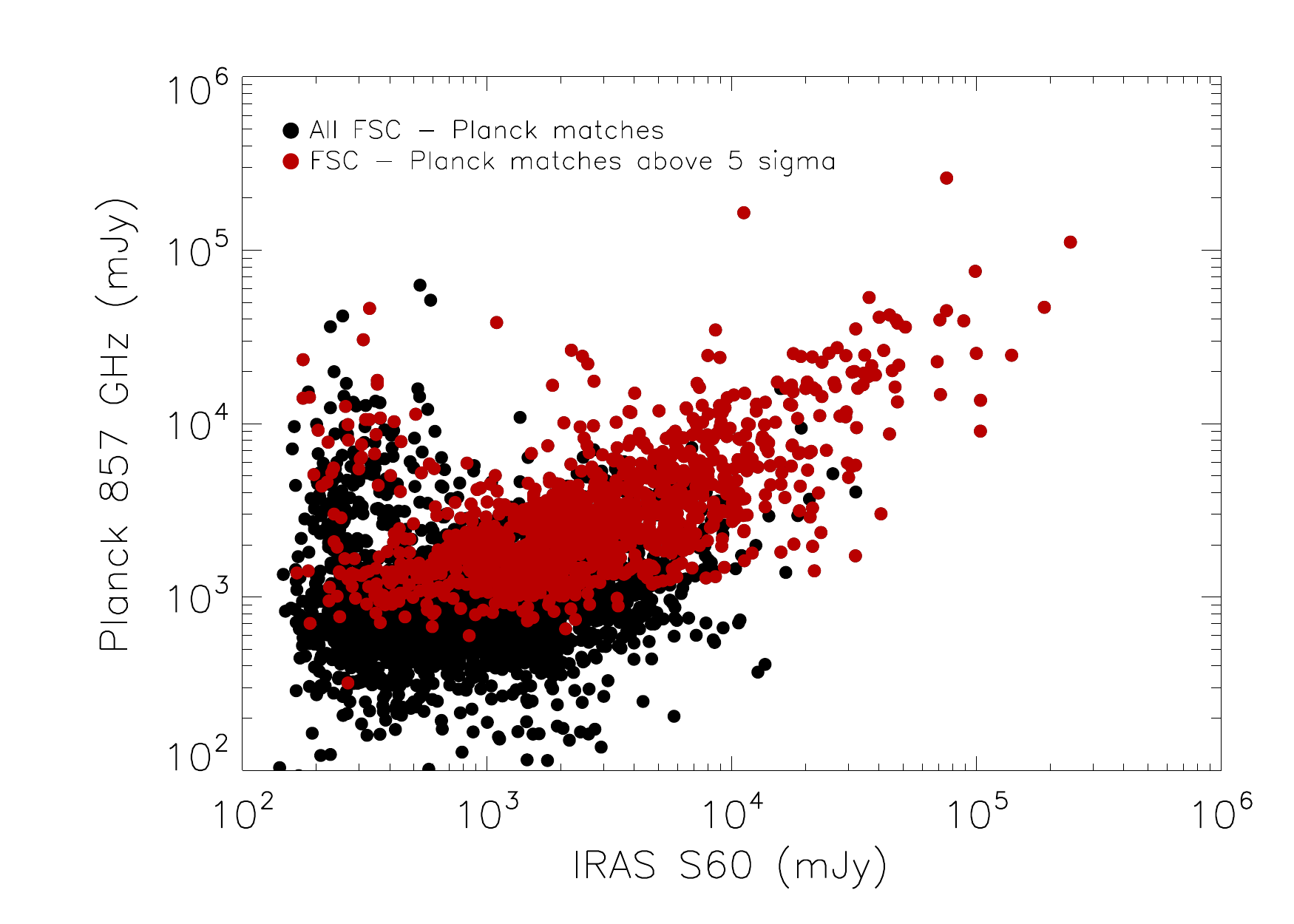}
\includegraphics[height=2.5in,width=3.4in]{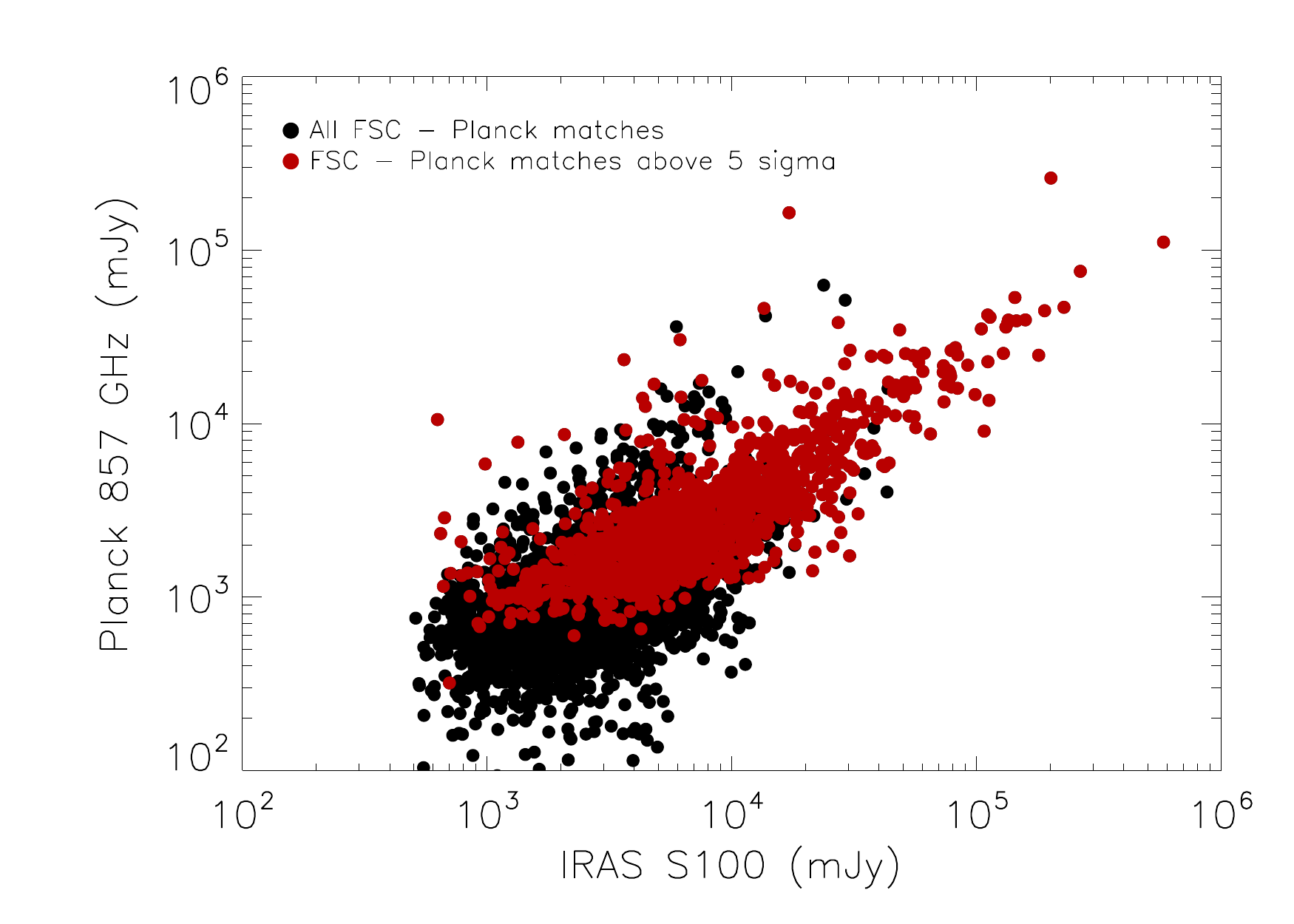}
\caption{IRAS 60\ $\micron$ (top panel) or 100\ $\micron$ (bottom panel) flux density versus Planck 857 GHz (350\ $\micron$) flux density for FSC - Planck matched sources. A generally good correlation between Planck and IRAS fluxes can be seen.}
\label{fig:iras_vs_planck}
\end{figure}

The first public release (PR1) of the Planck Catalogue of Compact Sources (PCCS; Planck Collaboration XXVIII 2013) consists of nine single-frequency catalogues of compact sources over the entire sky, covering the frequency range from 30 to 857 GHz (10561 to 350\ $\micron$). We cross-match the Planck PR1 catalogue at 857 GHz with the IIFSCz catalogue using a matching radius of 3\arcmin. The full width at half maximum (FWHM) of the Planck beam at 857 GHz is 4.33\arcmin. The 857 GHz catalogue also includes aperture photometry at 217 (1382\ $\micron$), 353 (850\ $\micron$) and 545 GHz (550\ $\micron$).  In total, there are 4,136 matched sources. Only about 30 Planck sources have more than one FSC sources within 3\arcmin. The histogram of positional differences between IRAS and Planck positions peaks at  40\arcsec. The comparison between the Planck 857 GHz flux densities and the IRAS 100\ $\micron$ or 60\ $\micron$ flux densities exhibits a good general correlation as shown in Fig.~\ref{fig:iras_vs_planck}. 



There are 1,200 IIFSCz sources with a Planck detection better than 5$\sigma$ in at least one of the 350, 550, 850 or 1382\ $\micron$ bands (and 2,364 sources with a planck detection better than 3$\sigma$). These are the most reliable Planck point-sources at $|b| > 20^o$, since for Planck sources unassociated with an IRAS extragalactic source there is a strong possibility that the source is due to cirrus.

\subsection{Spectroscopic redshift compilation}
\label{sec:redshiftCompilation}

The SDSS spectroscopic DR10 survey has covered a total unique area of 9274 deg$^2$  (Ahn et al. 2013). We get 11,749  redshift by cross-matching IRAS sources with WISE positions and SDSS objects. 
The 2MASS Redshift Survey (2MRS; Huchra et al. 2012) ultimately aims to determine the redshifts of all galaxies in the 2MASS XSC to a magnitude of K=12.2 mag and to within 5 deg of the Galactic plane. The second phase of 2MRS is now complete, providing an all-sky survey of 45,000 galaxies to a limiting magnitude of K=11.75 mag.  We retrieve 9, 264 redshifts by cross-matching IRAS sources with WISE positions and 2MASS sources. An NED all-sky query of FSC sources at $|b|>20^\circ$ with available redshift and S60 $>$ 0.1 Jy returned 64,219 objects.  9,523 new redshifts (not included in SDSS DR10 or 2MRS) are obtained. We also managed to get 4,557 spectroscopic redshifts from the PSC Redshift Survey (PSCz; Saunders et al. 2000),  the 6dF Galaxy Survey, and  the FSS redshift survey (FSSz; Oliver, PhD thesis). To summarise, we have obtained 33, 956 spectroscopic redshifts from SDSS DR10, 2MRS, NED, FSSz, PSCz and 6dF, which comprises $56\%$ of our base catalogue. 

Table 1 shows a breakdown of the multi-wavelength photometric data (from GALEX, SDSS, 2MASS, WISE, IRAS, Planck and NVSS) for the RIFSCz. Table 2 shows a summary of spectroscopic redshift data (from SDSS, 2MRS, NED, FSSz, PSCz and 6dF) for the RIFSCz.




\begin{table}
\caption{Summary of the multi-wavelength photometric data for the revised IRAS-FSC Redshift Catalogue. The statistics shown for the GALEX and Planck data require SNR $>3$. The statistics for the IRAS data require flux density quality $>=2$.}\label{zSource}
\begin{tabular}[pos]{ll}
\hline
Source          & Number   \\
\hline
GALEX NUV, FUV &  31,537/25,074\\
SDSS u,g,r,i,z  & 21,503 \\
2MASS  PSC J, H, K       & 45,892 \\
WISE  W1, W2, W3, W4              & 48,603/48,603/48,591/48,588  \\
IRAS 12, 25, 100\ $\micron$ & 4,476/9,606/30,942 \\
Planck 217, 353, 545, 857 GHz &  150/616/1,152/2,275\\
NVSS  1.4 GHz                        & 23,703   \\
\hline
\end{tabular}
\end{table}

\begin{table}
\caption{Summary of the spectroscopic redshift data for the revised IRAS-FSC Redshift Catalogue. Note that the number shown for a given spectroscopic survey or database does not exclude objects which are found elsewhere.}\label{zSource}
\begin{tabular}[pos]{lll}
\hline
Source of $z_{\rm spec}$         & Number   & Fraction \\
\hline
SDSS DR10  & 11,749    &$19\%$\\
2MRS                          & 9,264   &$15\%$\\
NED                          & 26,953    &$45\%$\\
IRAS FSSz                    & 568      &$0.9\%$\\
IRAS PSCz                    & 1,972     &$3.3\%$\\
6dF Galaxy Survey            & 2,794     &$4.6\%$\\
\hline
Total                        &33,956    &$56\%$\\
\hline
\end{tabular}
\end{table}

\section{PHOTOMETRIC REDSHIFT}

We apply a template-fitting method that has been used to construct the SWIRE Photometric
Redshift Catalogue (Rowan-Robinson et al. 2008 and references therein) to the revised IIFSCz galaxies with either optical or near-infrared photometry. Initially we use only a single pass through the data and a resolution of 0.002 in $\log_{10}(1+z)$. We use six galaxy templates (E, Sab, Sbc, Scd, Sdm and starburst) and three QSO templates, as in Rowan-Robinson et al. (2008). 


Our starting point is the revised IIFSCz sample of 60,303 sources associated with GALEX, SDSS, 2MASS, WISE, Planck and NVSS, and with obvious stars and cirrus sources removed.  33,956 sources now have spectroscopic redshifts through a compilation of several major spectroscopic surveys and the NED database.  We estimate photometric redshifts for the remainder using the available 0.361 - 4.6\ $\micron$ photometric data. We use SDSS model magnitudes and 2MASS extended fluxes, where available.  

\begin{figure}
\includegraphics[height=3.1in,width=3.4in]{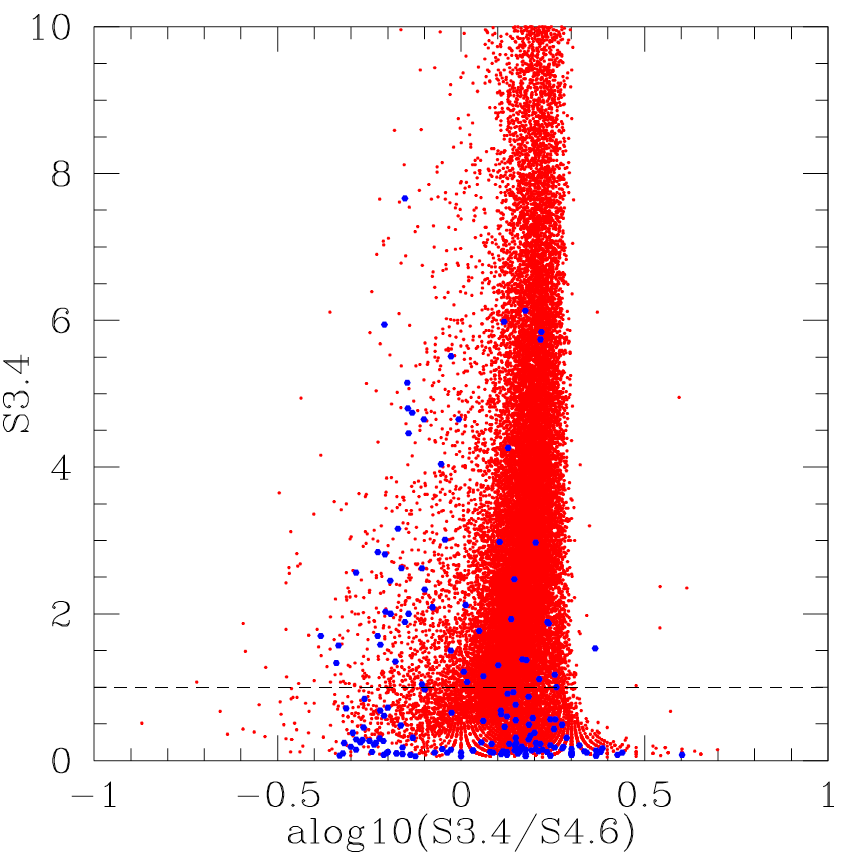}
\caption{The 3.4\ $\micron$ flux density (in mJy) S3.4 versus the 3.4-to-4.6\ $\micron$ flux density ratio $\log_{10} S3.4/S4.6$, exhibiting increased scatter at S3.4 $<$ 1 mJy (below the dashed line). Red dots are galaxies and blue dots are quasars in the RIFSCz.}
\label{plot3446galqso6f}
\end{figure}

Fig.~\ref{plot3446galqso6f} shows the WISE 3.4$/$4.6\ $\micron$ colour versus 3.4\ $\micron$ flux density S3.4.  There is clearly increased scatter at the faintest 3.4\ $\micron$ fluxes. Thus, we only use WISE 3.4 and 4.6\ $\micron$ fluxes in the photometric redshift estimation if S3.4 $>$ 1 mJy and S4.6 $>$ 1 mJy. We also require S12 $>$ 1 mJy, S24 $>$ 5 mJy (corresponding roughly to the 5-$\sigma$ WISE limits).

\subsection{Aperture corrections}

\begin{figure*}
\includegraphics[height=3.1in,width=3.4in]{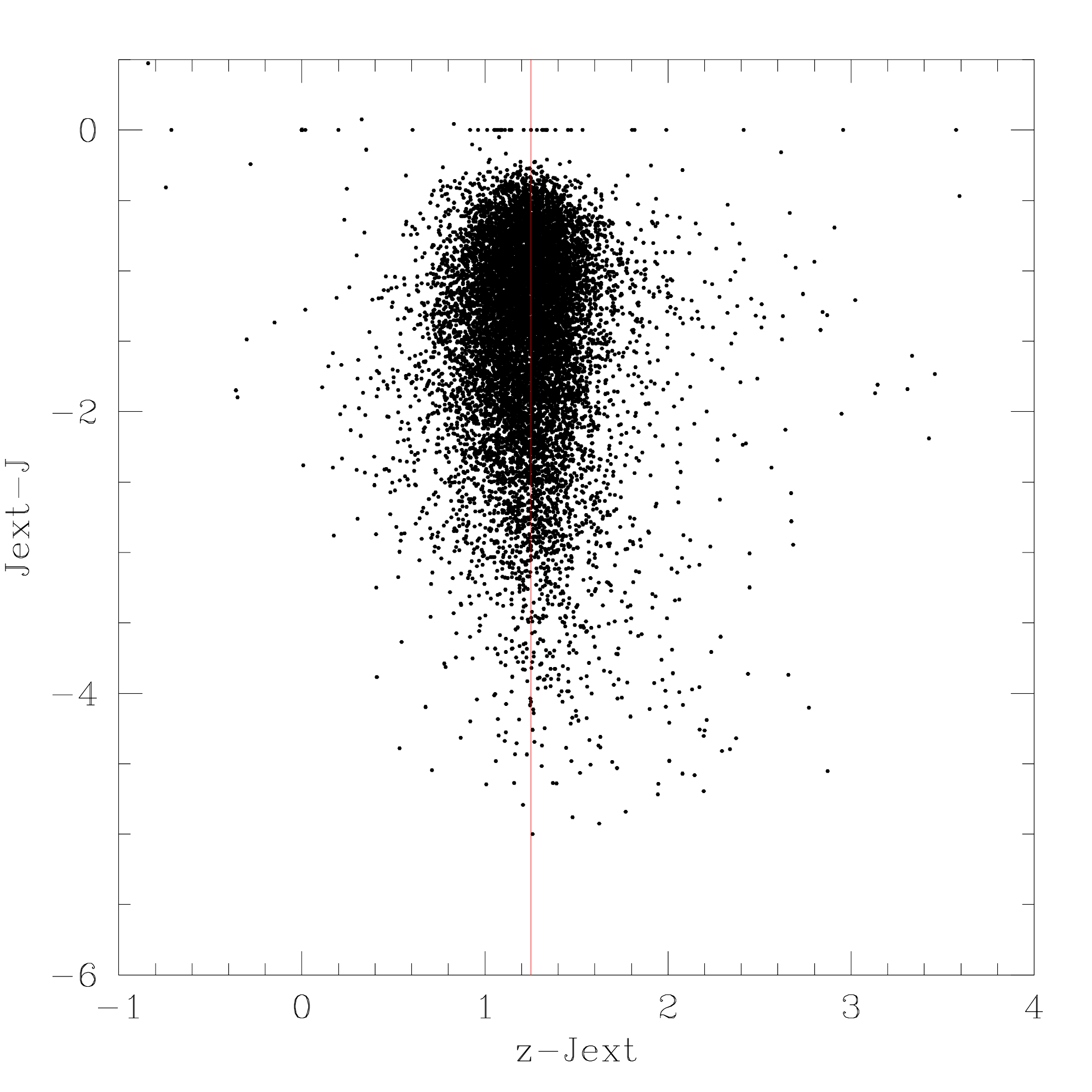}
\includegraphics[height=3.1in,width=3.4in]{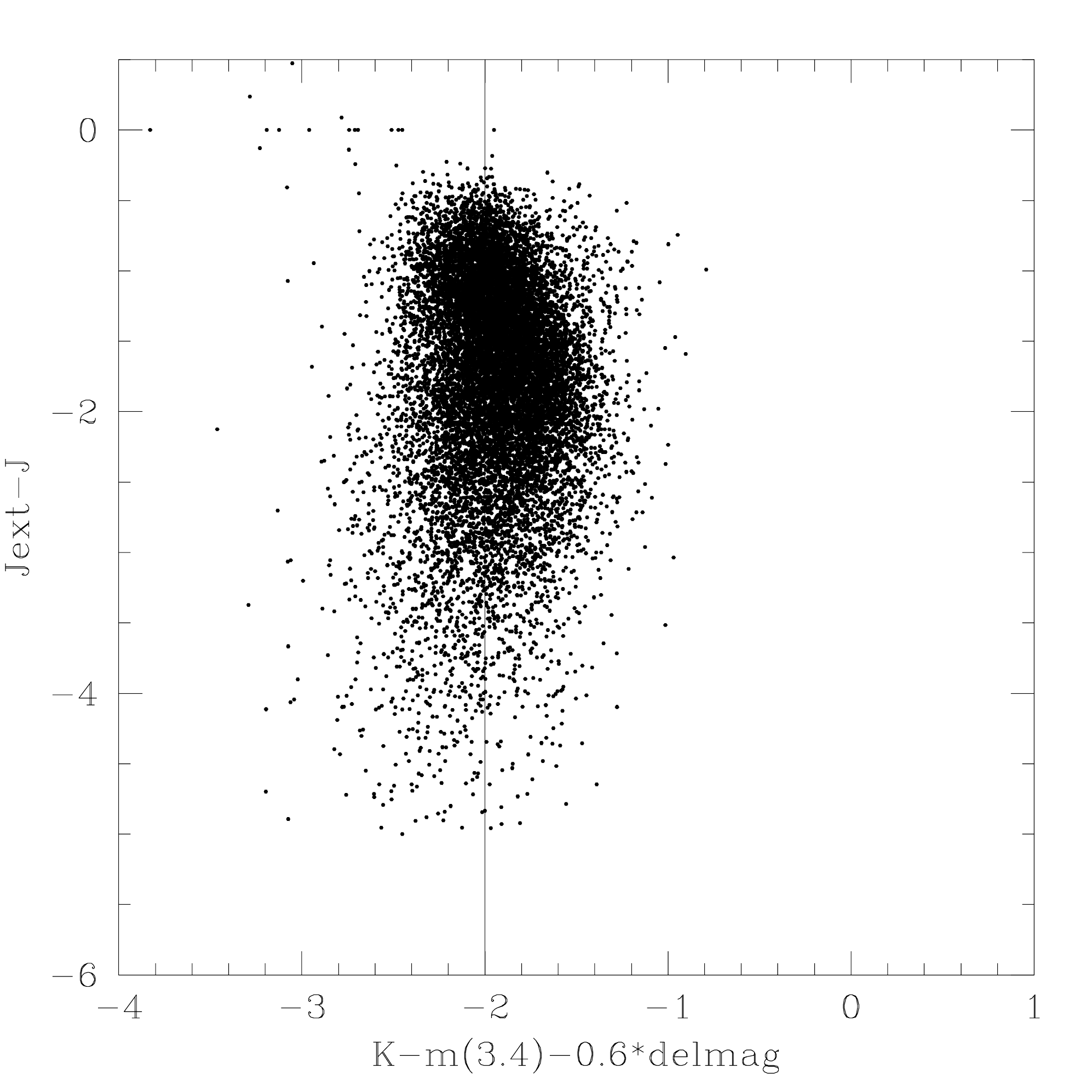}
\caption{LH: $J_{\rm ext} - J$ versus ($z-J_{\rm ext}$), illustrating that the SDSS model magnitudes and 2MASS extended magnitudes have well-matched aperture corrections. RH: $J_{\rm ext} - J$ versus ($K_{\rm ext}-m(3.4)-0.6\times(J_{\rm ext} - J$)), illustrating that the WISE 3.4\ $\micron$ magnitudes need an aperture correction = 0.6 $\times$ the 2MASS J-band aperture correction.}
 \label{plotdelmag}
\end{figure*}

The left panel in Fig.~\ref{plotdelmag} shows the J-band aperture correction ($J_{\rm ext} - J$) versus ($z-J_{\rm ext}$), illustrating that  the SDSS model magnitudes and 2MASS extended magnitudes have well-matched aperture corrections. However a similar plot for  ($K_{\rm ext}-m(3.4)$) shows a strong dependence of colour on J-band aperture correction, suggesting that the WISE 3.4 and 4.6\ $\micron$ magnitudes need to be aperture corrected.  The right panel in Fig.~\ref{plotdelmag} shows ($J_{\rm ext} - J$) versus ($K_{\rm ext}-m(3.4)-0.6\times(J_{\rm ext} - J$)), illustrating that the WISE 3.4\ $\micron$ magnitudes need an aperture correction = 0.6 $\times$ the 2MASS J-band aperture  correction, delmag = $J_{\rm ext} - J$. We have applied this same aperture correction to the WISE 4.6\ $\micron$ data. 

We have also explored whether the WISE 12 and 22\ $\micron$ fluxes need aperture correction, and found optimum results with
\begin{equation}
\log_{10} S12_{\rm corr} = \log_{10} S12 - 0.16\times {\rm delmag}
\end{equation}
and
\begin{equation}
\log_{10} S22_{\rm corr} = \log_{10} S22 - 0.10\times {\rm delmag}
\end{equation}
where delmag is the J-band aperture correction, $J_{\rm ext} - J$.

\begin{figure*}
\includegraphics[height=3.1in,width=3.4in]{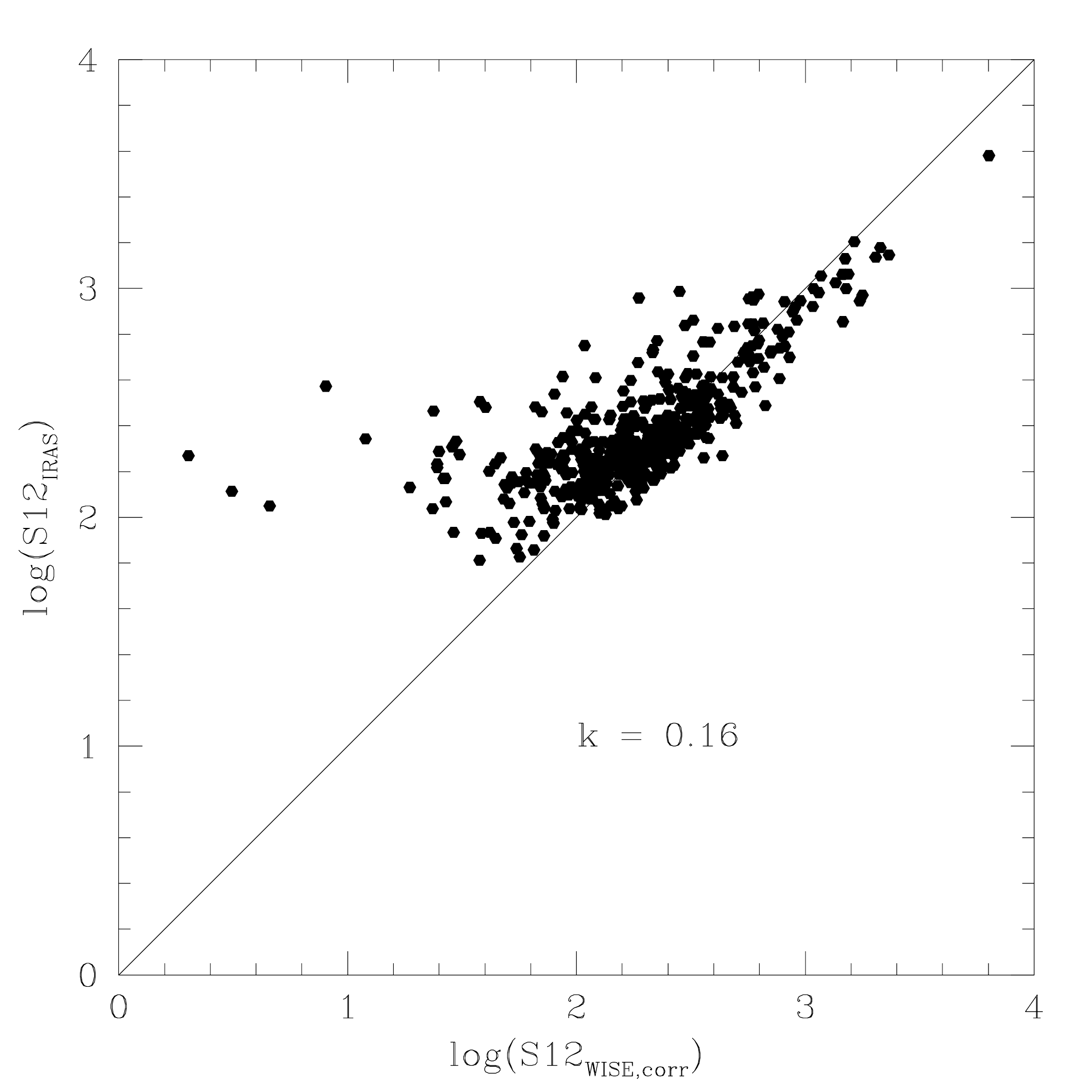}
\includegraphics[height=3.1in,width=3.4in]{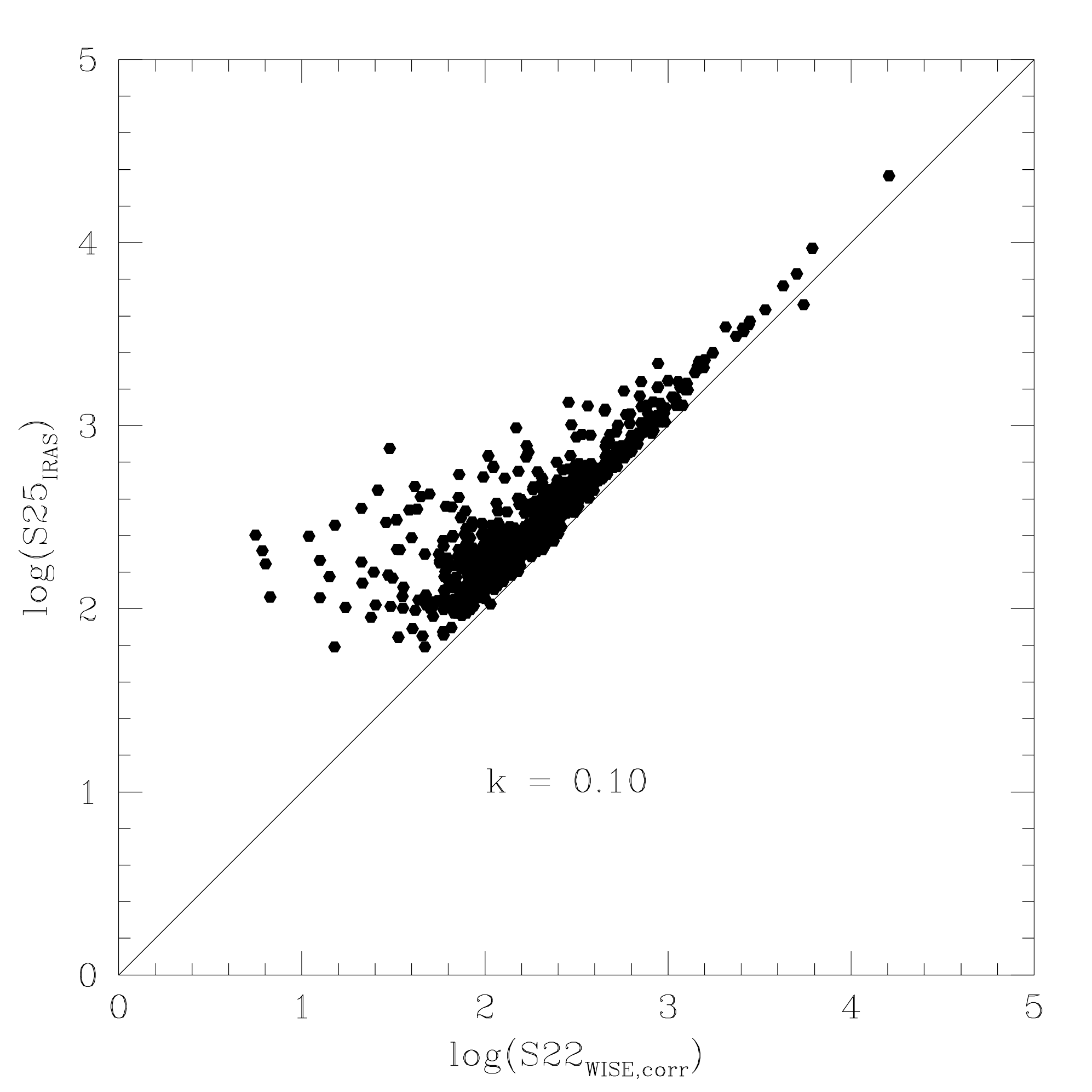}
\caption{Left panel: IRAS 12\ $\micron$ fluxes versus aperture corrected WISE 12\ $\micron$ fluxes (with $S12>1.0$ mJy). The k value in the plot is the coefficient in the aperture correction in the WISE 12\ $\micron$ band as in Eq. (5). Right panel: IRAS 25\ $\micron$ fluxes versus aperture corrected WISE 22\ $\micron$ fluxes (with $S22>5$ mJy). The k value in the plot is the coefficient in the aperture correction in the WISE 22\ $\micron$ band as in Eq. (6). There is a good agreement in both panels except for a few sources where the WISE fluxes seem to be underestimated. The alternative explanation that the IRAS fluxes of these sources are overestimated  does not fit the SED modelling.}
\label{IRASvsWISE}
\end{figure*}

The left panel in Fig.~\ref{IRASvsWISE} shows the IRAS 12 $\micron$ flux versus the aperture corrected WISE 12 $\micron$ flux. The right panel in Fig.~\ref{IRASvsWISE} shows the IRAS 25\ $\micron$ flux versus the aperture corrected WISE 22\ $\micron$ flux. The agreement is good in both cases, except for a tail of sources where the WISE fluxes appear to be underestimated.  The alternative  interpretation, that the IRAS fluxes are overestimated does not fit in well with SED modelling of these sources.

\subsection{Stellarity}

To distinguish reliably between galaxies and QSOs without generating excessive aliasing, a  measure of stellarity is required.  SDSS sources with sdsstyp = ``STAR" are given a stellar flag, as are NED sources with nedtp = ``QSO". In addition we have assumed that sources with $J_{\rm ext} - J > -0.3$ are stellar.

\subsection{Identification of AGN dust tori}

A feature of the codes developed for estimating photometric redshifts for Spitzer galaxies (Rowan-Robinson et al 2005, 2008, 2012) is the use of 3.6 and 4.5\ $\micron$ data to constrain redshift.  This is very effective in reducing the number of catastrophic outliers.  However it becomes important to identify those galaxies with significant AGN dust tori, whether QSOs or Seyferts, since the dust torus can dominate over starlight (or AGN continuum) at wavelengths down to 1\ $\micron$.  

\begin{figure}
\includegraphics[height=3.2in,width=3.4in]{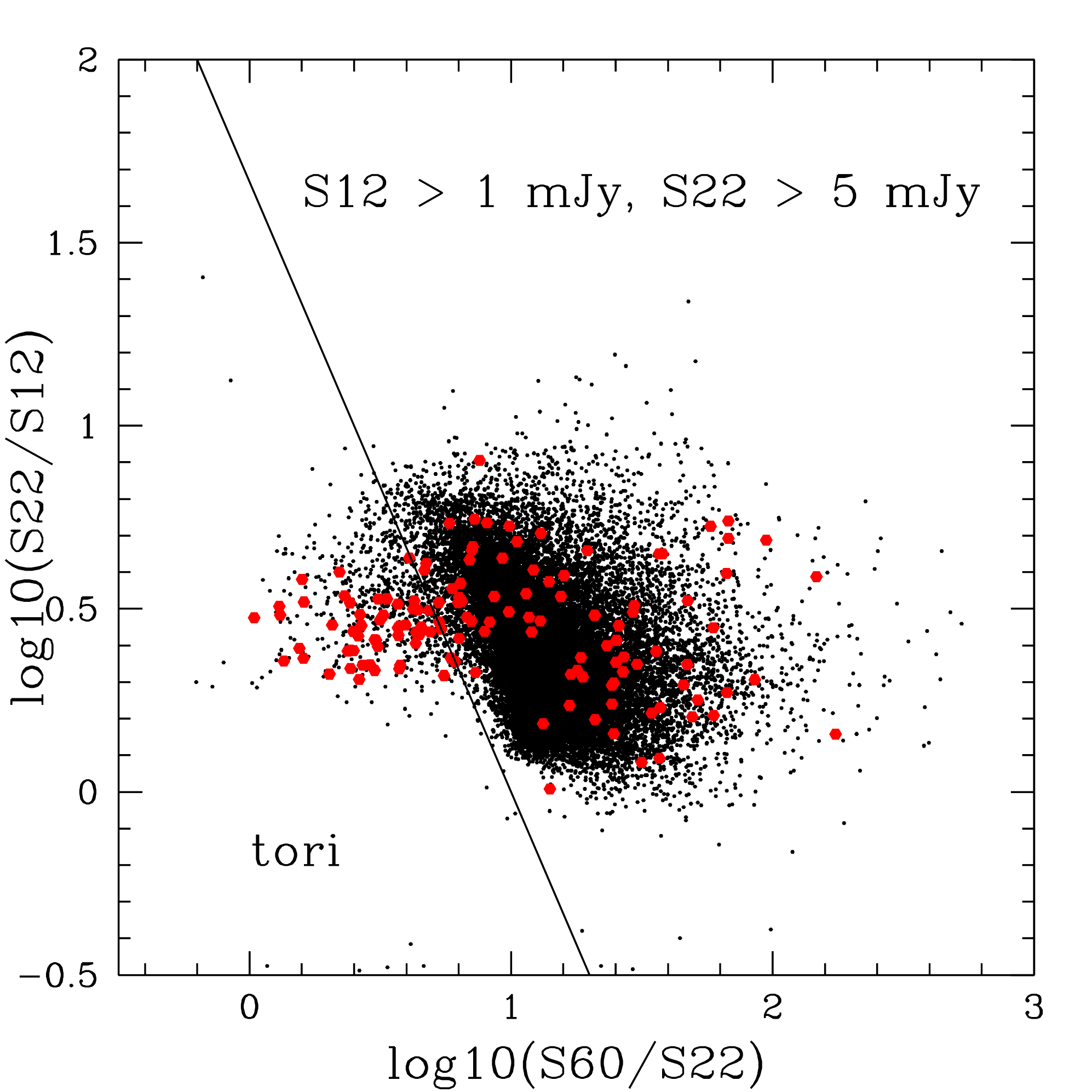}
\caption{The 22-to-12\ $\micron$ flux ratio, $S22/S12$, versus the 60-to-22\ $\micron$ flux ratio, $S60/S22$, with QSOs shown in red. Objects to the left of the sloping line are assumed to have significant dust tori and thus go through a second photometric redshift pass.}
\label{plot122260fssgalqso4}
\end{figure}

If we want to use the WISE 3.4 and 4.6\ $\micron$ fluxes  (and the J, H, and K$_s$ fluxes) in the redshift solution, we have to find a way of identifying the AGN dust tori.  We have chosen to use a key diagnostic diagram introduced by Rowan-Robinson and Crawford (1969), the 12-25-60\ $\micron$ colour-colour diagram.  Fig.~\ref{plot122260fssgalqso4} shows a plot of $\log_{10} S60/S22$ versus $\log_{10} S22/S12$ for the RIFSCz sample, with 12 and 22 $\micron$ fluxes taken from WISE and the 60\ $\micron$ flux taken from IRAS.   Sources to the left of the sloping line in Fig.~\ref{plot122260fssgalqso4} are assumed to have significant dust tori and these (together with all sources identified as QSOs on the first pass) go through a second photometric redshift pass in  which, following Rowan-Robinson et al (2012), each galaxy or QSO template has added to it ten different possible amplitudes of an AGN dust torus.  WISE 3.4 and 4.6\ $\micron$ data are  used only in the second pass, for the sources suspected to contain AGN dust tori.  For galaxies without AGN dust tori, the 3.4 and 4.6 $\micron$ did not improve the redshift solution, perhaps because of inadequate aperture correction of extended sources.

\subsection{Performance}

\begin{figure}
\includegraphics[height=3.2in,width=3.4in]{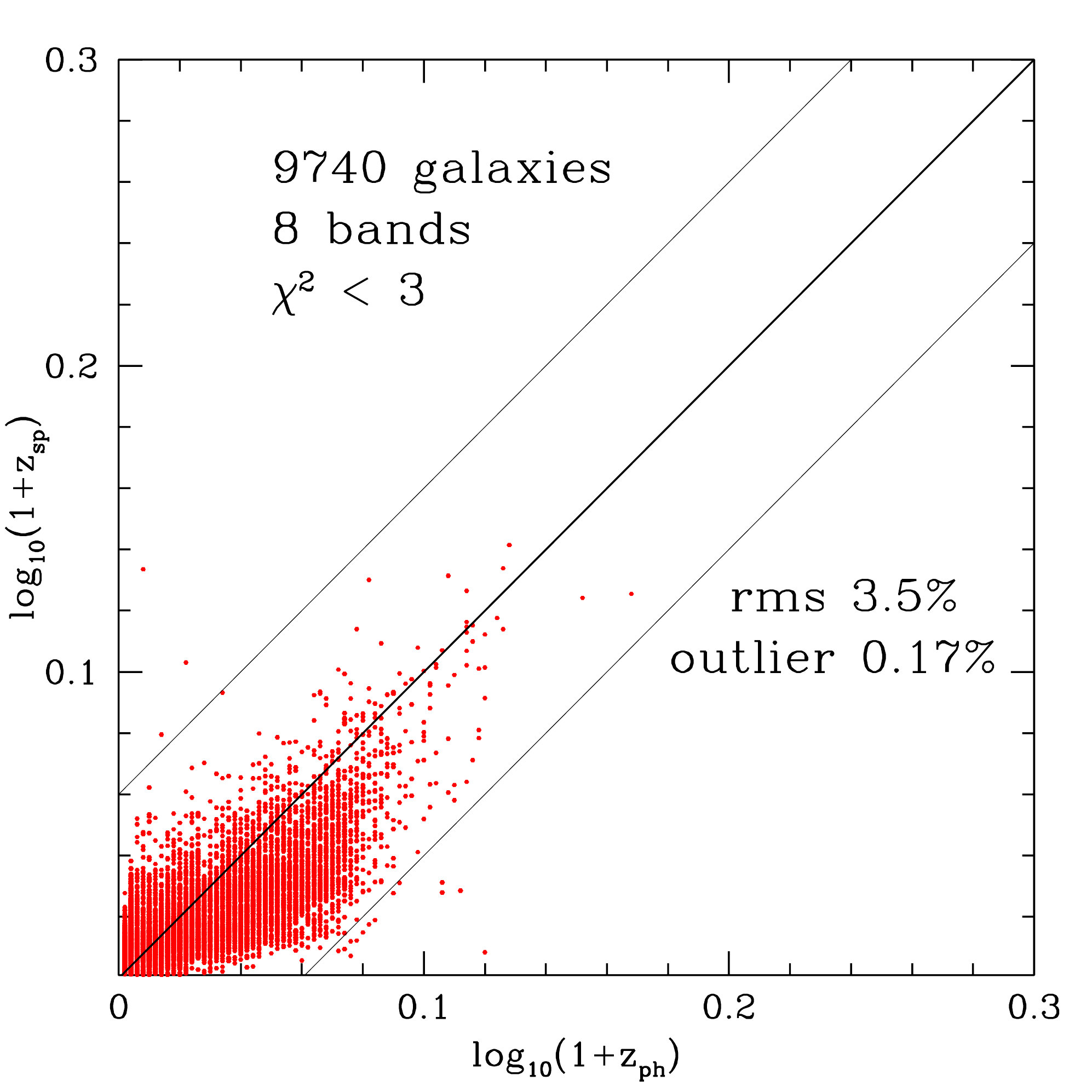}
\includegraphics[height=3.2in,width=3.4in]{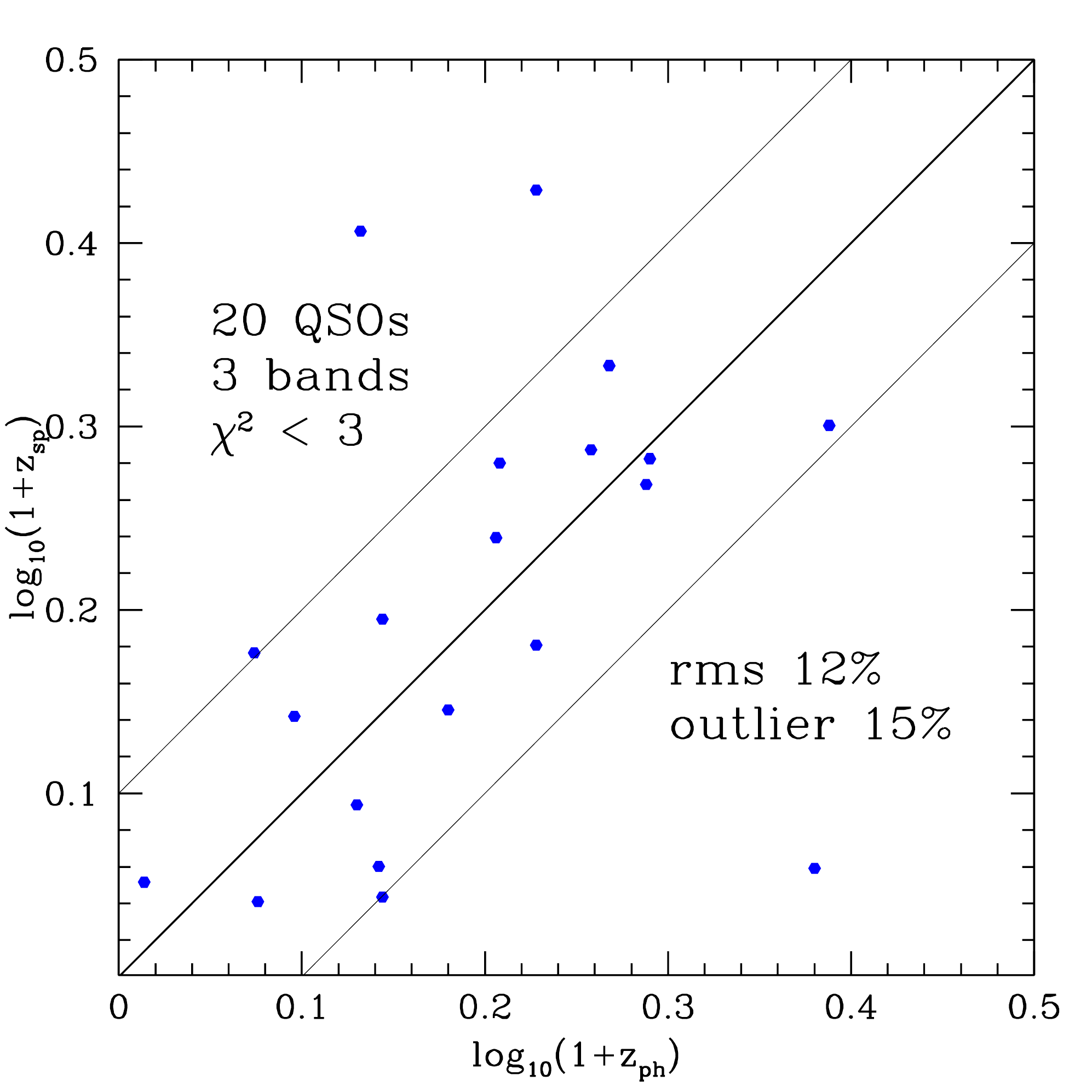}
\caption{Top: Spectroscopic versus photometric redshift for all galaxies (red) and QSOs (blue), with 8 photometric bands and reduced $\chi^2 < 3$. The thick diagonal line corresponds to $1+z_{\rm phot} = 1+ z_{\rm spec}$. Galaxies outside the two thin diagonal lines are outliers with $|\log_{10} (1+z_{\rm phot}) - \log_{10} (1+z_{\rm spec})|>0.06$ (i.e. $|\Delta z/(1+z_{\rm spec})| > 0.15$). Bottom: Spectroscopic versus photometric redshift for all QSOs with at least 3 photometric bands and reduced $\chi^2 < 3$. The thick diagonal line corresponds to $1+z_{\rm phot} = 1+ z_{\rm spec}$. QSOs outside the two thin diagonal lines are outliers with $|\log_{10} (1+z_{\rm phot}) - \log_{10} (1+z_{\rm spec})|>0.1$ (i.e. $|\Delta z/(1+z_{\rm spec})| > 0.26$).}
\label{zplotfssfinal8irb8err3}
\end{figure}

The top panel in Fig.~\ref{zplotfssfinal8irb8err3} shows a comparison of photometric and spectroscopic redshifts for sources with at least 8 photometric bands and with reduced $\chi^2 < 3$. The percentage of catastrophic outliers, i.e. ($1+z_{\rm phot}$) differs from ($1+z_{\rm spec}$) by more than 15$\%$, is 0.17, 3.4 and 3.5 $\%$ for sources with 8,  5 and 3 photometric bands, respectively.  The rms accuracy, after  exclusion of these outliers, is 3.5, 4.0 and 3.8 $\%$ for sources with 8, 5 and 3 photometric bands,  respectively.  These represent a good photometric redshift performance (cf Rowan-Robinson et al 2008, 2013). The bottom panel in Fig.~\ref{zplotfssfinal8irb8err3} shows a comparison of spectroscopic and photometric redshifts for QSOs with at least 3 photometric bands and reduced $\chi^2 <3$. The approximately power-law form of the optical continua of QSOs leads to some redshift aliasing. To summarise, we have 33,956 spectroscopic redshifts, 15,406 photometric redshifts, 37 sources with photometry  but no successful redshift due to inconsistencies among the photometry data, and 10,941 sources with no spectroscopic or photometric redshift.

\begin{figure*}
\includegraphics[height=3.2in,width=3.45in]{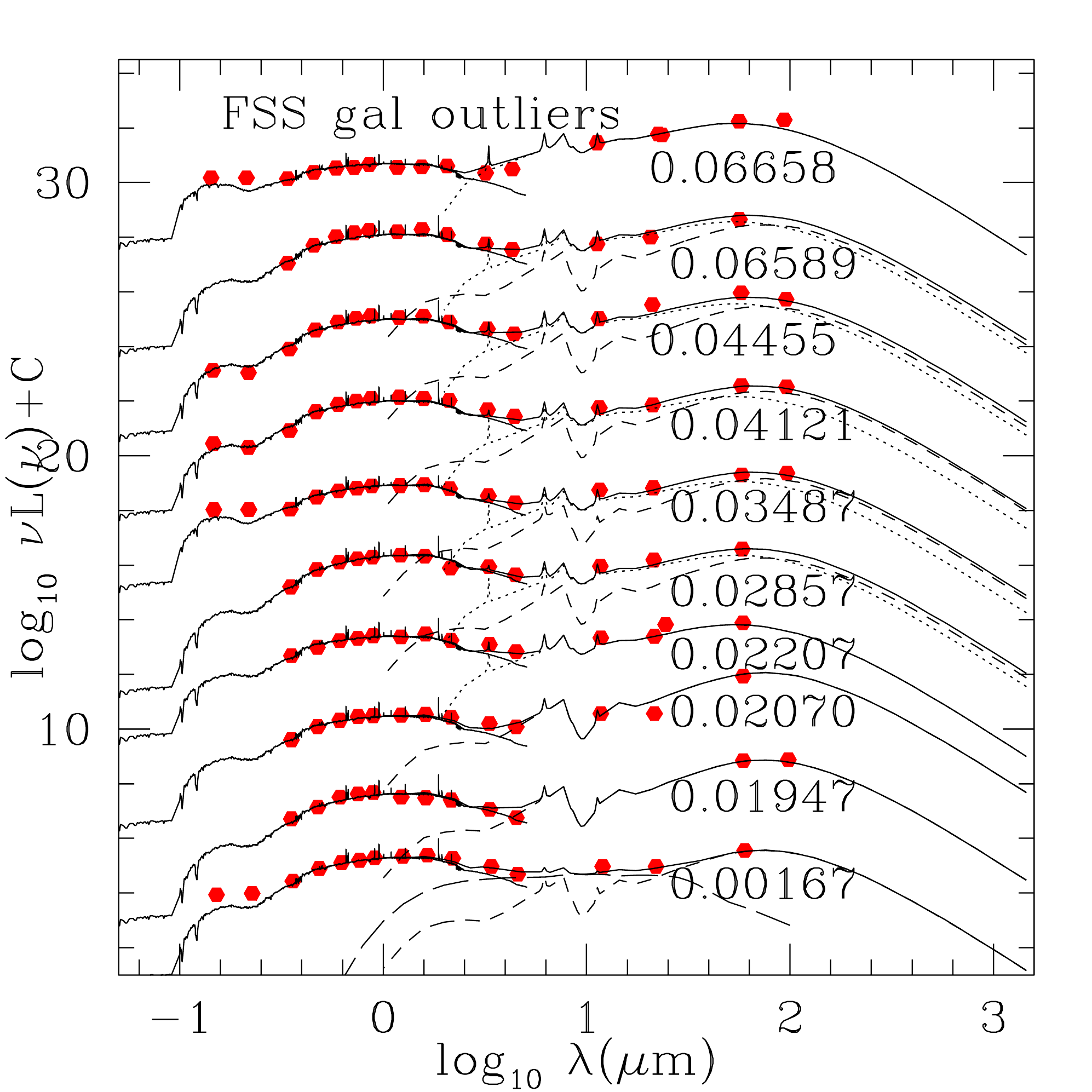}
\includegraphics[height=3.2in,width=3.45in]{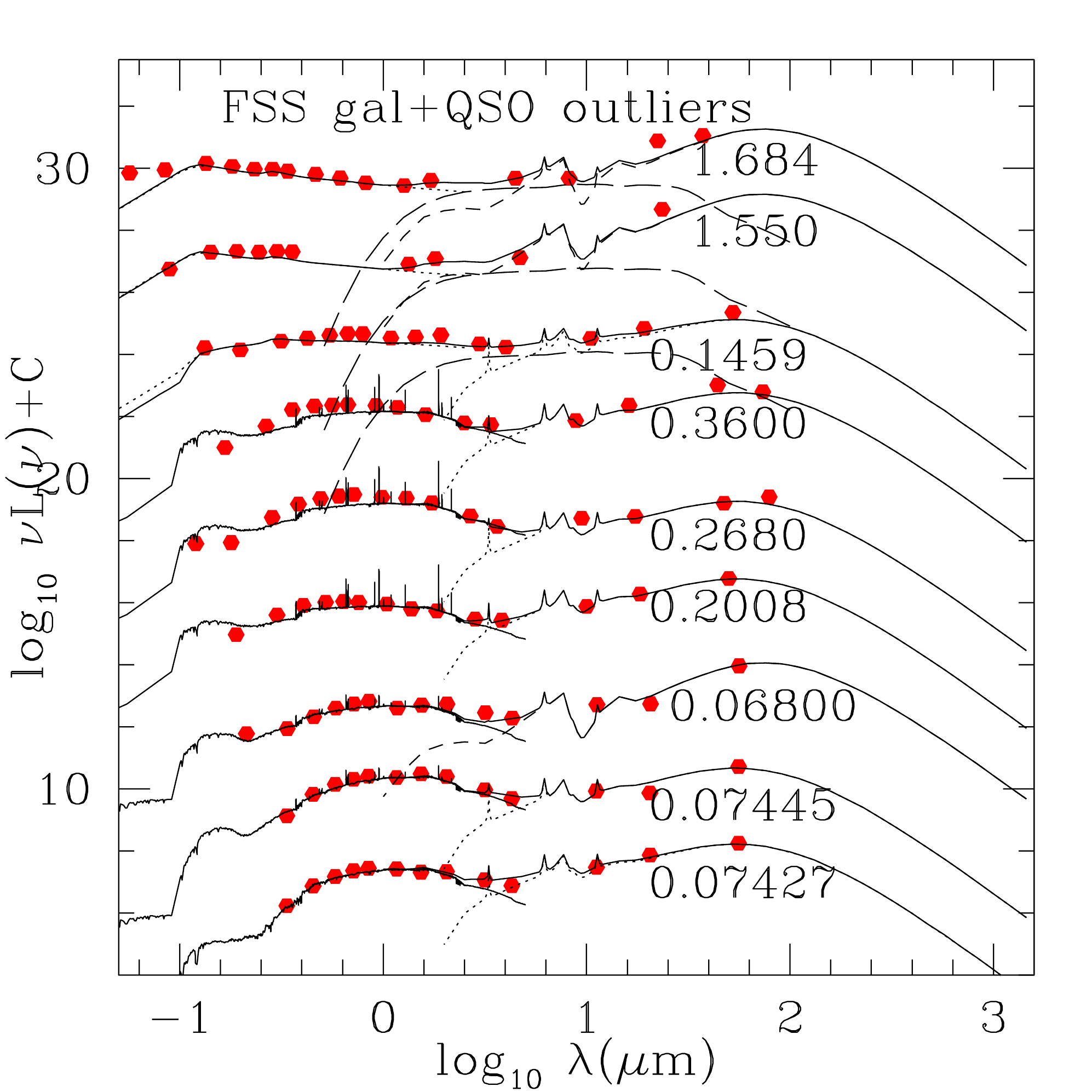}
\caption{The SEDs of galaxies and QSOs which are outliers in the spectroscopic redshift versus photometric redshift plot (Fig.~\ref{zplotfssfinal8irb8err3}), offset vertically for clarity. Dotted curves correspond to the M82 starburst template, short-dashed curves correspond to the Arp 220 starburst template, and the long-dashed curves correspond to the AGN dust torus template. The starlight component and total emission are both shown as continuous curves. The spectroscopic redshifts, as indicated near each object, seem plausible in all cases. The photometric redshifts could be pulled away from the correct value by small mismatches in the SDSS and 2MASS aperture corrections, and/or by aliasing caused by dust extinction.}
\label{probfssnewg45sed}
\end{figure*}

Fig.~\ref{probfssnewg45sed} shows SEDs of galaxies (and one QSO) which are outliers with $|\log_{10}(1+z_{\rm phot}) - \log_{10}(1+z_{\rm spec})| > 0.06$ (i.e. $|\Delta z/(1+z_{\rm spec})| > 0.15$) in Fig.~\ref{zplotfssfinal8irb8err3}.  Galex data have been used in the SED plots, where available, but not in the photometric redshift solution. The spectroscopic redshifts, as indicated near each object,  look plausible in all cases.  The photometric redshift appears to be pulled away from the correct value by small mismatches in the SDSS and 2MASS aperture corrections, and by aliasing caused by dust extinction. Where the aperture corrected SDSS and 2MASS fluxes are highly discrepant we have used only 2MASS fluxes in the redshift solution.
 
\section{Infrared template fits}

Mid- and far-infrared data from IRAS and WISE (12, 22, 25, 60 and 100\ $\micron$) and sub-mm data from Planck (350, 550 and 850\ $\micron$) were fitted initially, following the methodology of Rowan-Robinson et al (2005, 2008) and as in Wang \& Rowan-Robinson (2009a), with a combination of four infrared templates (cirrus, M82 and Arp220 starbursts, or AGN dust torus). However while the four standard infrared templates work well for many sources, the 350 - 850\ $\micron$ fluxes often require the presence of colder dust than is incorporated into our four basic templates.  The two new templates used here are taken from the range of optically thin interstellar medium (``cirrus") templates developed by  Rowan-Robinson (1992) and Efstathiou \& Rowan-Robinson (2003). The key parameter determining the temperature of the  dust grains is the intensity of the radiation field, which we can characterise by the ratio of intensity of  radiation field to the local interstellar radiation field, $\psi$.  The standard cirrus template corresponds to $\psi$ = 5, and this is the value used by Rowan-Robinson (1992) to fit the central regions of our Galaxy.  $\psi$ = 1 corresponds to the interstellar radiation field in the vicinity of the Sun.  We also find that some galaxies need a much lower intensity radiation field than this, with $\psi$ = 0.1.  The corresponding grain temperatures in the dust model of Rowan-Robinson (1992) for the two new template are in the ranges 14.5 - 19.7 K and 9.8 - 13.4 K respectively.  Full details of the templates used are given at \url{http://astro.ic.ac.uk/mrr/swirephotzcat/templates/readme}.

Following the analysis of Rowan-Robinson \& Efstathiou (2009) of the Spoon et al (2008) IRS diagnostic diagram, we have also introduced a young starburst (ysb) template, more extreme than the Arp220 template.  This was found to be necessary by Rowan-Robinson et al (2010) in their analysis of the SEDs of a preliminary sample of 68 Lockman-SPIRE sources. Another template considered, following Rowan-Robinson \& Efstathiou (2009), was an old starburst template, but we found that the SED for this was too similar to a pure cirrus template to be a worthwhile addition.

\begin{figure}
\includegraphics[height=3.2in,width=3.4in]{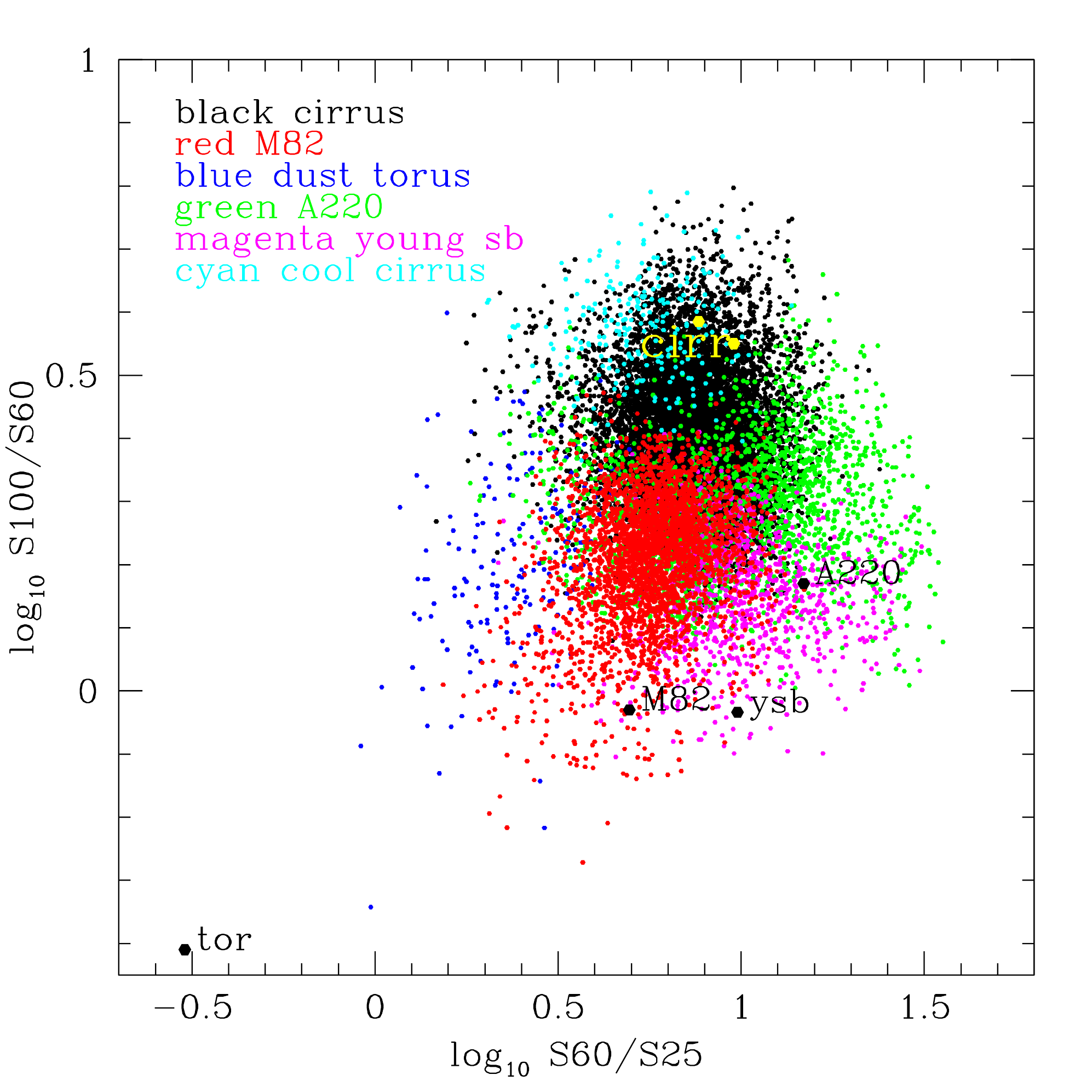}
\caption{The 25 (or 22)-60-100\ $\micron$ colour-colour diagram for RIFSCz sources, colour-coded by the infrared template making the dominant contribution to the infrared luminosity.  Location of individual templates in this plot are shown. It is clear that standard cirrus ($\psi=5$) or cool cirrus ($\psi=1$) dominated sources have colder 60-100\ $\micron$ colours, while starburst or AGN dominated sources have warmer 60-100\ $\micron$ colours.}
\label{plot256010026ir11ht}
\end{figure}

Our automatic SED fitting code uses two cirrus templates ($\psi$ = 5, 1), one of three starburst templates (M82, Arp220, ysb), and an AGN dust torus template.  These infrared SED templates provide a good fit to the mid- and far-infrared, and sub-mm, data for most sources. Fig.~\ref{plot256010026ir11ht}  shows the 25 (or 22) - 60 - 100\ $\micron$ colour-colour diagram, colour-coded by the infrared template making the dominant contribution to the infrared luminosity.  Sources whose SEDs are dominated by standard cirrus ($\psi=5$) or cool cirrus ($\psi=1$) have colder 60-100\ $\micron$ colours, while starburst or AGN dominated sources have warmer 60-100\ $\micron$ colours.

\begin{figure}
\includegraphics[height=3.2in,width=3.4in]{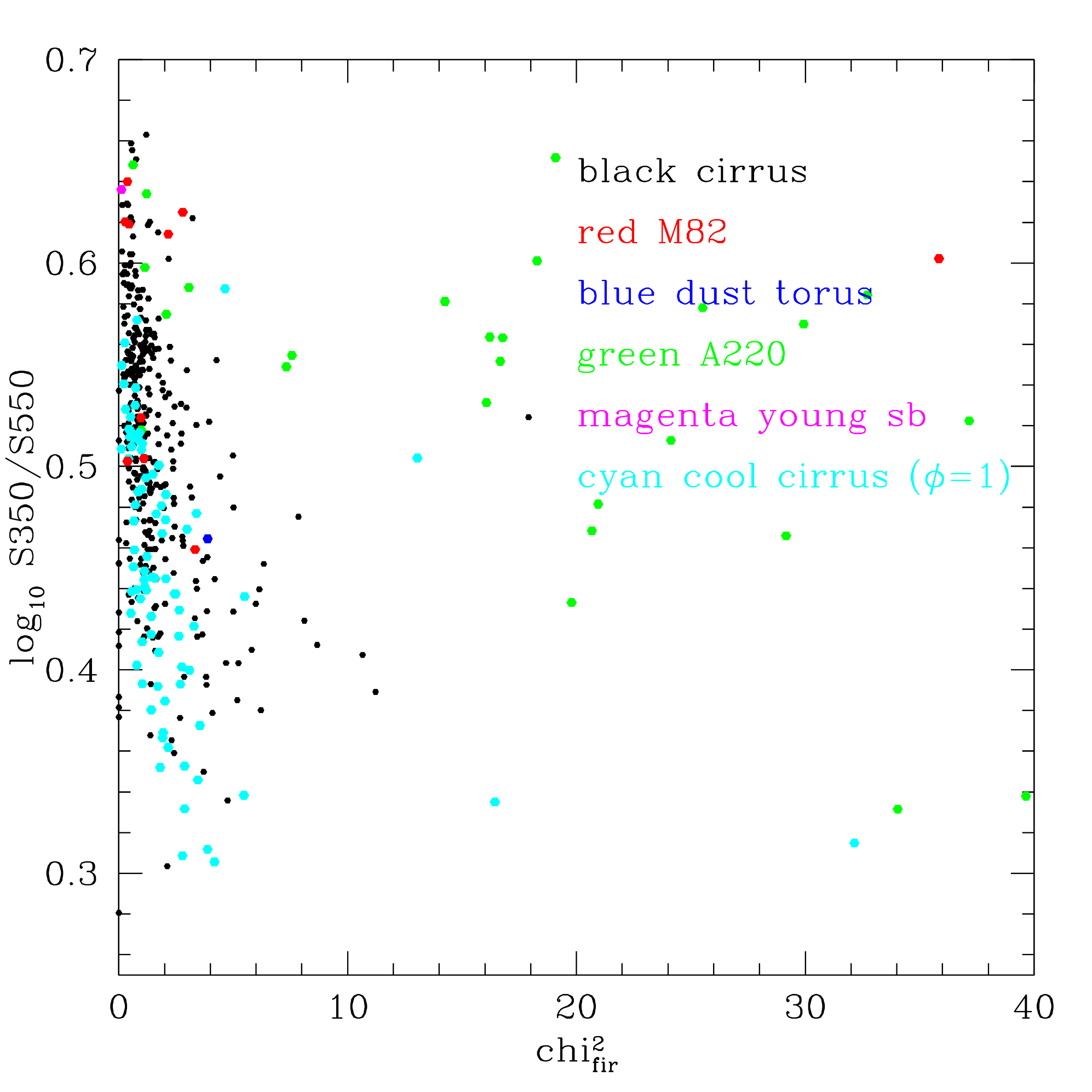}
\caption{The Planck 350$/$550\ $\micron$ flux density ratio versus the reduced $\chi^2$ for the infrared and sub-millimetre fit. It is clear that the reduced $\chi^2$ is good for most of the sources and especially for warmer sources.}
\label{plot350550errir6ir11ht}
\end{figure}

Fig.~\ref{plot350550errir6ir11ht} shows the Plankc 350$/$500\ $\micron$ flux-ratio versus the reduced $\chi^2$ for the infrared and sub-mm fit. This illustrates that the reduced $\chi^2$ is good for most sources, and especially for warmer sources.  We have modelled the SEDs of all sources with sub-mm data for which $\chi^2 > 10$.  For many cases the poor $\chi^2$  is simply due to incorrect aperture corrections, especially for the WISE fluxes.  For very large galaxies which are in the IRAS Large Optical Galaxy Catalog (Rice et al 1988), or which have delmag $<$ -3, we do not use the WISE fluxes in the redshift or infrared template fit solutions.

\begin{figure*}
\includegraphics[height=3.4in,width=3.4in]{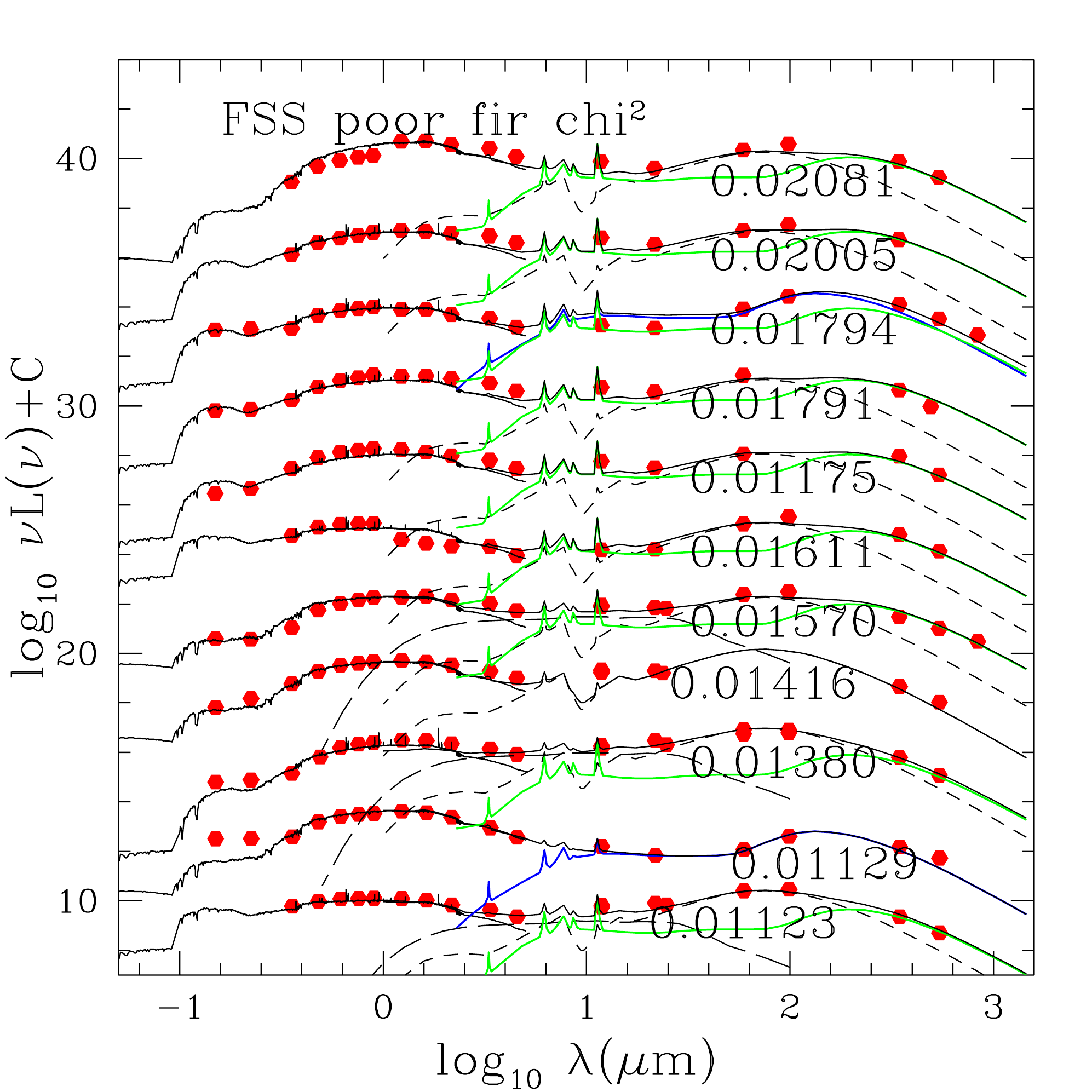}
\includegraphics[height=3.4in,width=3.4in]{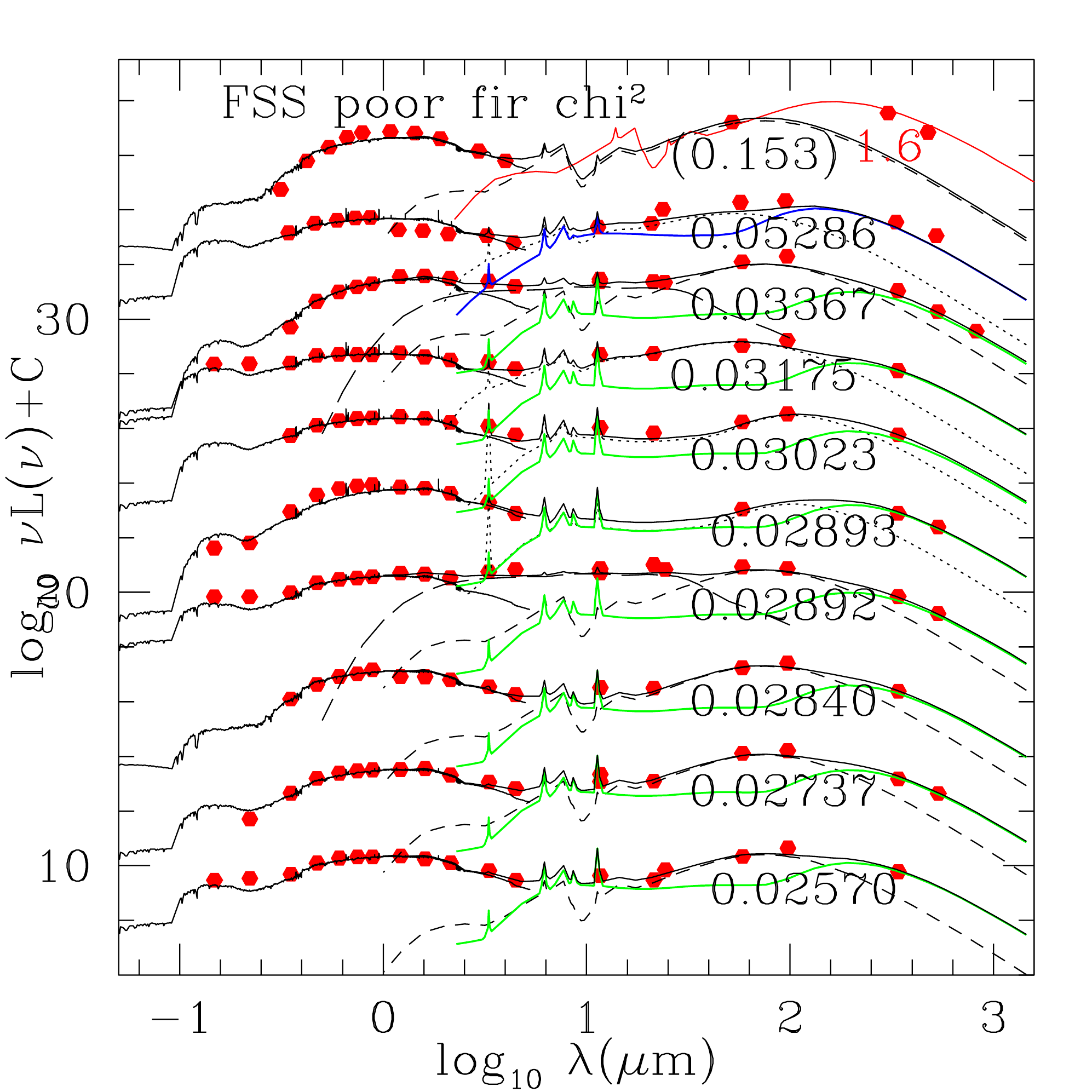}
\caption{SEDs for sources with good Planck 350\ $\micron$ detections and poor $\chi^2$ for the infrared and sub-millimetre template fit.  Blue line denotes the cool cirrus ($\psi$=1) template, green line denotes the cold cirrus ($\psi$=0.1) template, and red line denotes a candidate gravitational lens. For most sources, the fit can be improved by adding a cold cirrus template.}
\label{probfssnewir}
\end{figure*}

In most cases there is a need for the cold cirrus component ($\psi$ = 0.1), as found previously by Rowan-Robinson et al (2010) and Planck Collaboration XVI (2011).  Fig.~\ref{probfssnewir} shows SEDs for 21 sources with good 350\ $\micron$ detections, $z>0.01$, 8 optical and near-infrared photometric bands and $\chi^2_{fir} > 10$.  A further 2 (not shown) are QSOs (3C273, OJ287), which we do not expect to be fitted by our templates. For 19 of these sources the fit is improved by adding a cold cirrus component. One source (F01464+1954) is a candidate to be a lensed galaxy.


\subsection{The IRAS-Planck sources}

There are 1,200 RIFSCz sources which have been matched to a Planck detection better than $5\sigma$ in at least one of the 350, 550, 850 and 1382\ $\micron$ bands (and 2,386 sources with a planck detection $>3\sigma$ in at least one band). They are also of great interest because for these sources we have the strongest constraints on their far-infrared and sub-mm spectrum.  

\begin{figure*}
\includegraphics[height=3.2in,width=3.4in]{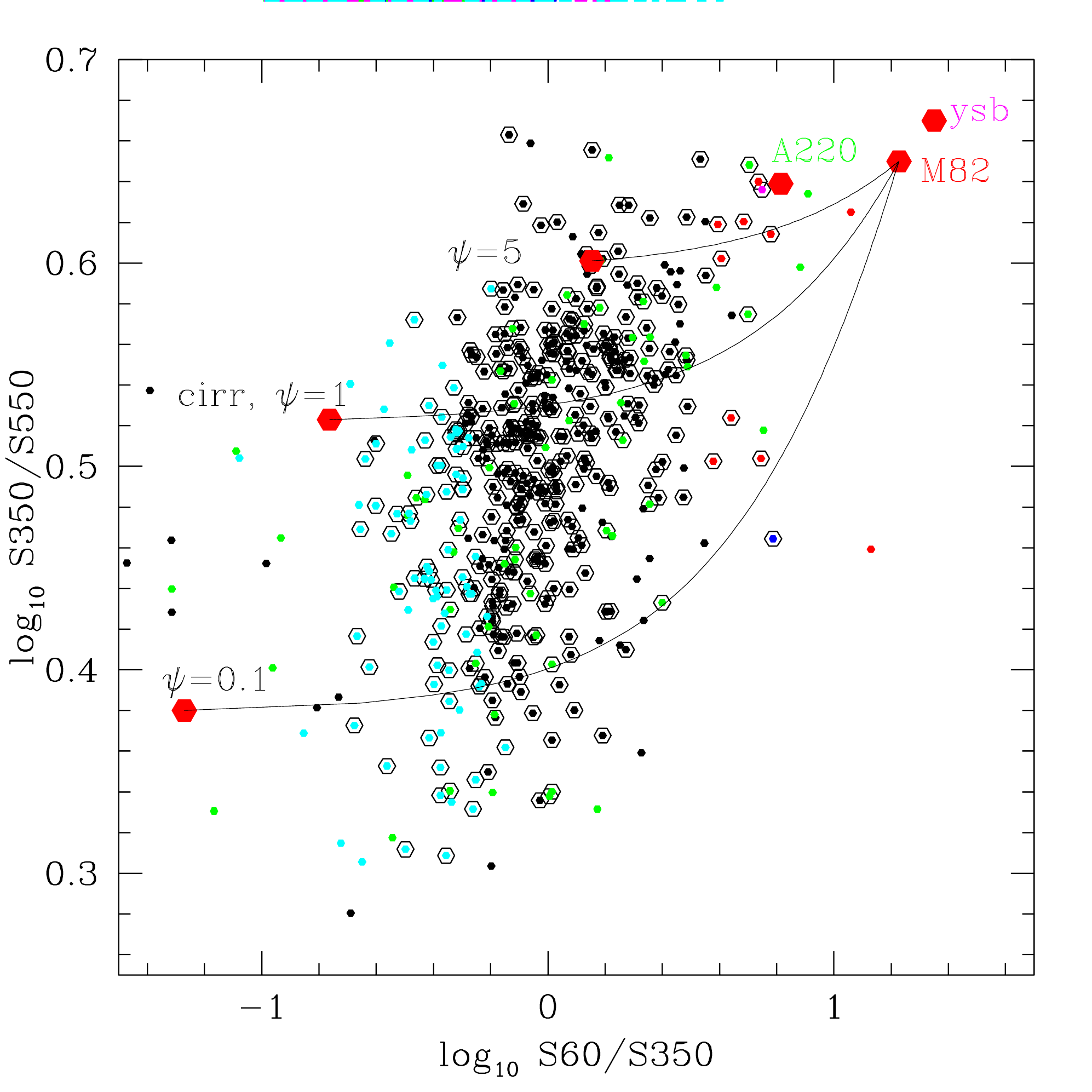}
\includegraphics[height=3.2in,width=3.4in]{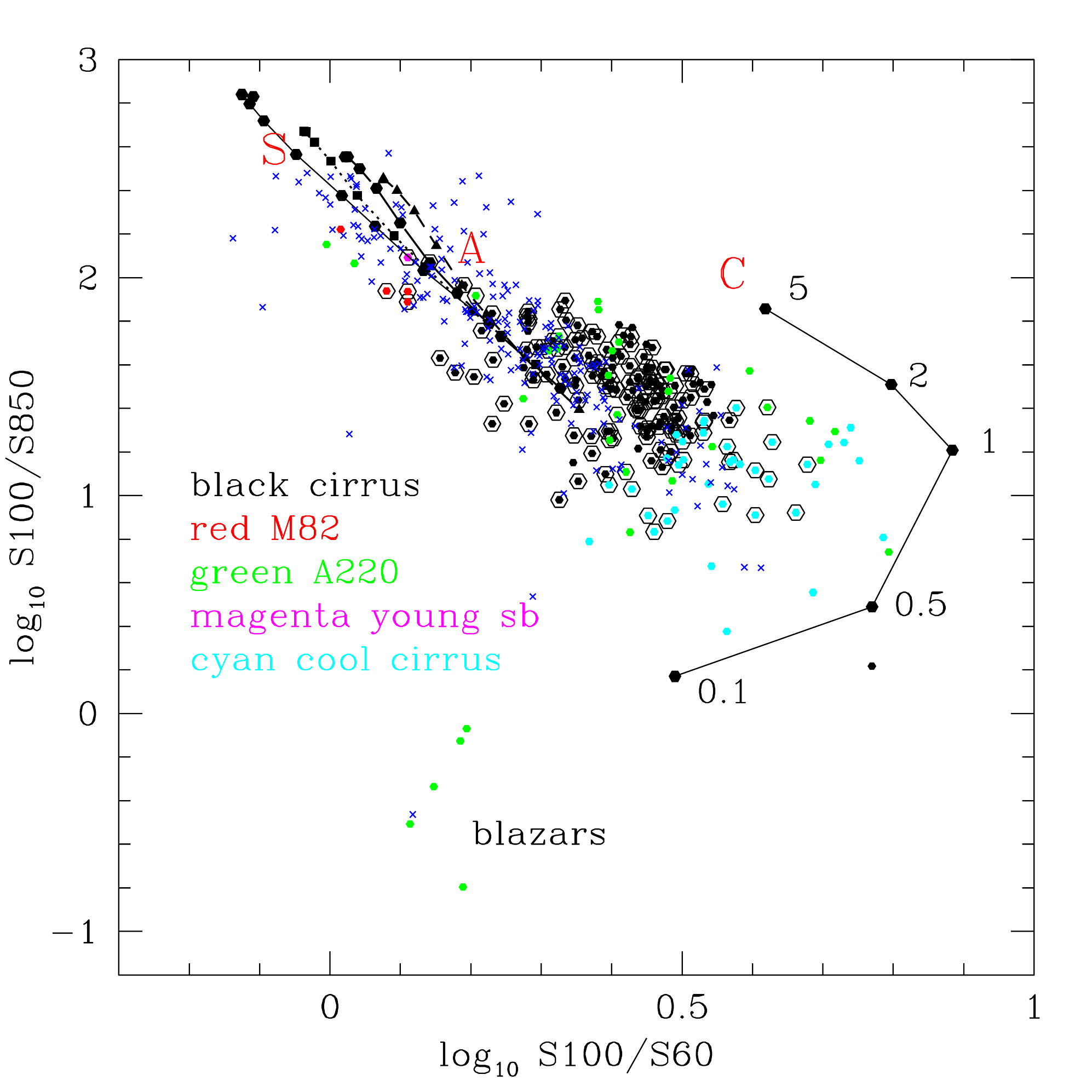}
\caption{Left panel: S350$/$S550 versus S60$/$S350 for IRAS-Planck matched sources, colour-coded by infrared template type. Galaxies from the IRAS Large Optical Galaxy Catalog (Rice et al 1988) are shown as open hexagons.  The locations of our infrared templates are indicated, and  mixture lines between an M82 starburst template and the three cirrus templates are shown. Right panel: S100$/$S850 versus S100$/$S60 for Planck-IRAS galaxies (filled circles) and for a compilation of IRAS sources with ground-based 850\ $\micron$ detections (blue crosses).
Sequences of starburst models from Efstathiou et al (2000) with optical depth (from the left) $\tau_{uv}$ = 50, 100, 150, 200: standard M82 and Arp220 starburst templates marked S and A.  A sequence of cirrus models C with $\psi$ = 0.1, 0.5, 1, 2, 5 is also shown.}
\label{plot601008506ir11h6pierre}
\end{figure*}

The left panel in Fig.~\ref{plot601008506ir11h6pierre} shows the 60 - 350 - 550\ $\micron$ colour-colour diagram (S350$/$S550 versus S60$/$S350)  for 5$\sigma$ Planck sources, with locations of our infrared model templates indicated. A further illustration of the need for a cold cirrus component is proved by this figure. The location of the different templates (at $z=0$) are indicated, together with a mixture line between an M82 starburst and the cold cirrus ($\psi$ = 0.1) template.



The right panel in Fig.~\ref{plot601008506ir11h6pierre} shows S100$/$S850 versus S100$/$S60 for RIFSCz galaxies and for a compilation of ground-based 850\ $\micron$ detections (P. Chanial personal communication), together with model sequences of starburst and cirrus templates.  The group of 6 sources at the bottom of the plot are known blazers.  This figure again illustrates the need for cooler cirrus templates to explain the observed sub-millimetre fluxes.

\subsection{Infrared luminosity, star-formation rate, stellar mass and dust mass}

\begin{figure}
\includegraphics[height=3.22in,width=3.42in]{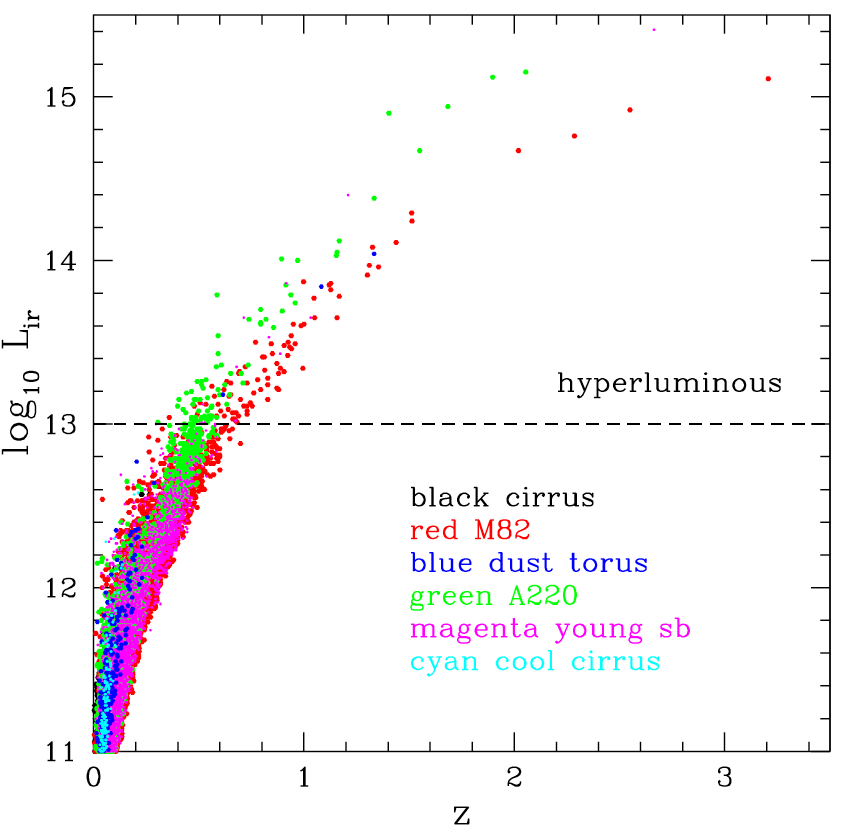}
\includegraphics[height=3.22in,width=3.45in]{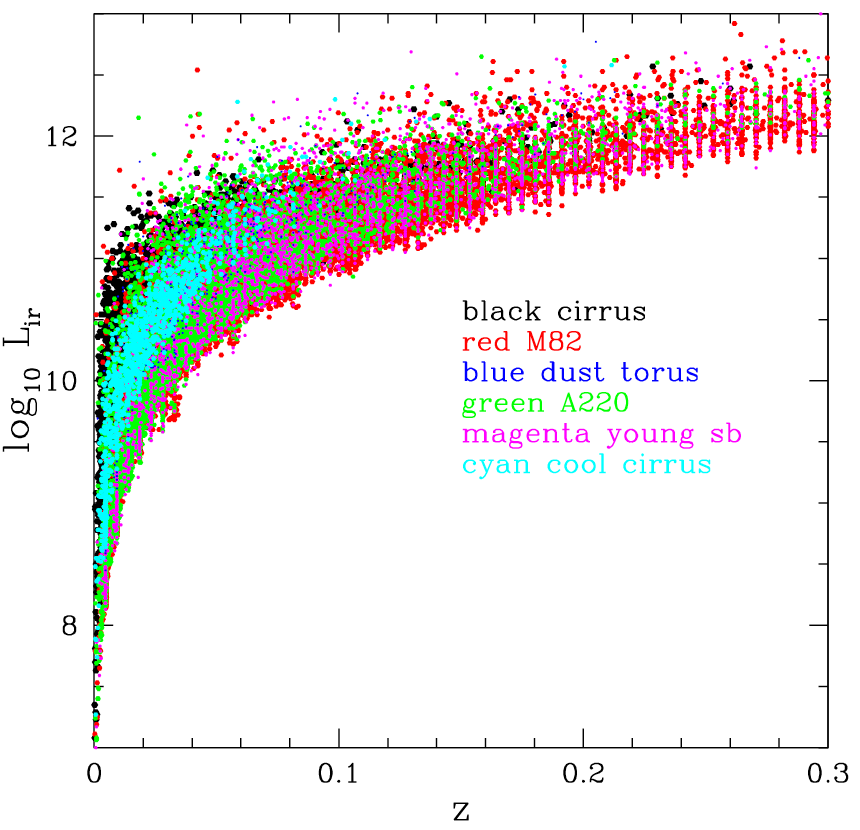}
\caption{Top: Infrared luminosity versus redshift for the RIFSCz sources, colour-coded by the infrared template making the dominant contribution to the infrared luminosity. Sources above the dashed line are hyper luminous infrared galaxies with infrared luminosity $>10^{13}L_{\odot}$, which are mostly either M82 type or A220 type starbursts. Bottom: Close-up for galaxies with redshift $<0.3$.} 
\label{plotirlumzfinal28ir}
\end{figure}

Fig.~\ref{plotirlumzfinal28ir} shows the infrared luminosity versus redshift for the  RIFSCz sources, colour-coded by infrared template type.  Where sources selected cirrus templates with $\log_{10} L_{fir} > 12.5$, or have $L_{\rm cirrus}>L_{\rm opt}$,  sources were put through a second pass with such unphysical cirrus luminosities not permitted.  Many sources found acceptable fits with other templates.  Those that had poor reduced $\chi^2$ in the infrared template fit may be a consequence of mis-association of the IRAS source with other catalogues, or of incorrect redshifts, or may be candidate of gravitationally lensed objects. 

178 galaxies are found to by hyper-luminous ($L_{fir} > 10^{13} L_{\odot}$) (cf Rowan-Robinson 2000), of which 61 have spectroscopic redshifts. This is very similar to the 179 found by Rowan-Robinson \& Wang (2010) in the earlier version of the IIFSCz Catalogue. It would be worthwhile to obtain redshifts for the remaining 118 candidate hyper-luminous objects.

\begin{figure}
\includegraphics[height=3.2in,width=3.45in]{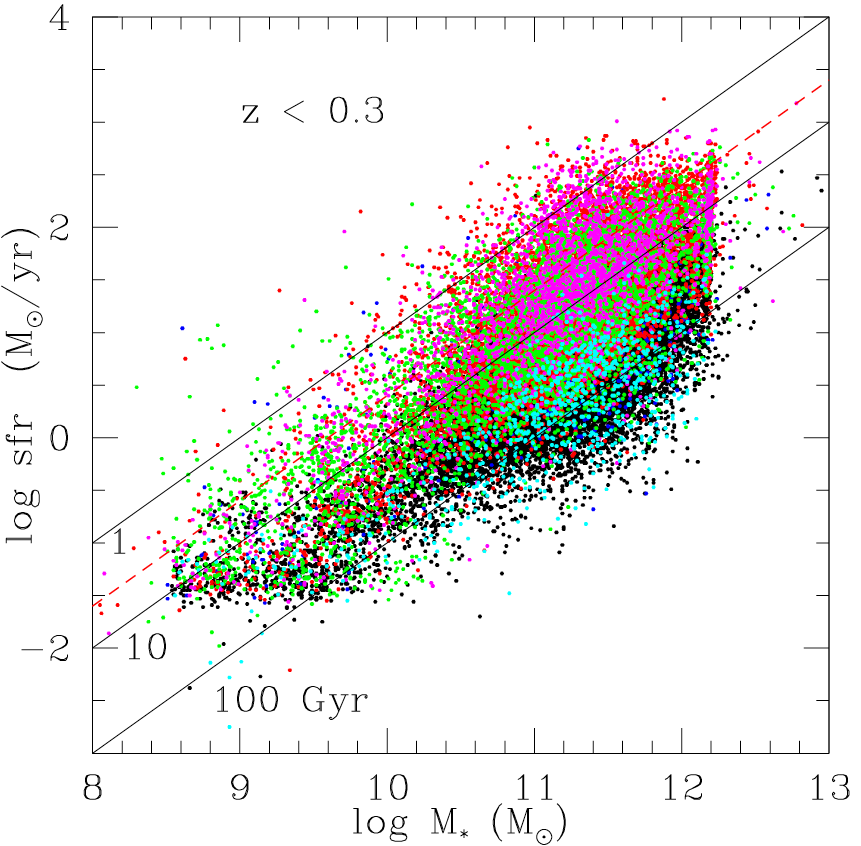}
\includegraphics[height=3.2in,width=3.45in]{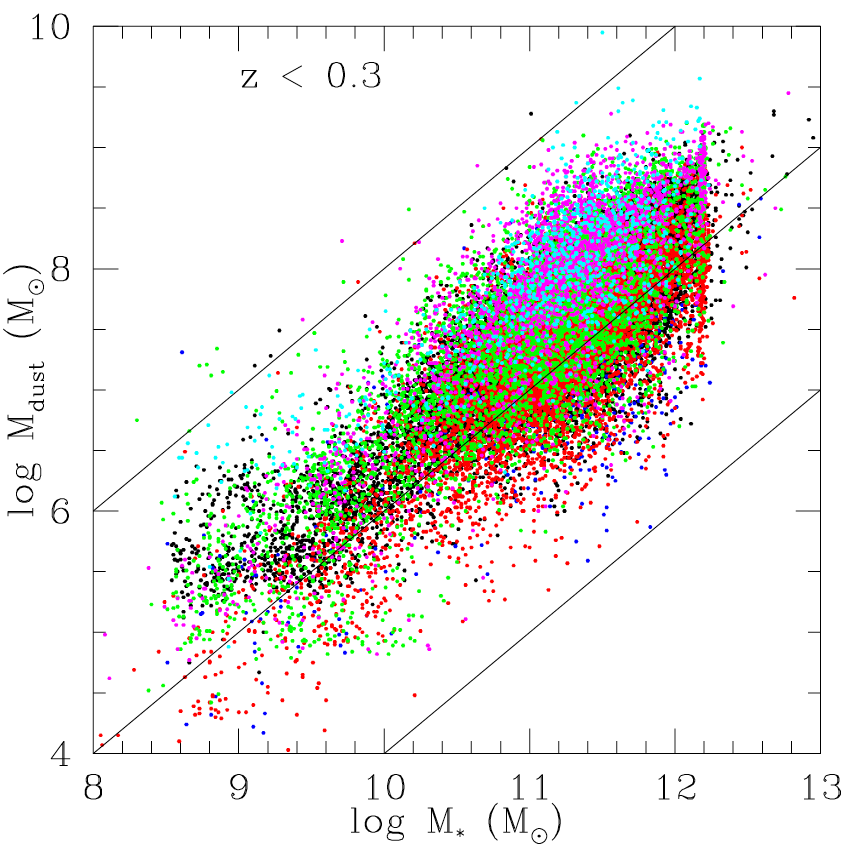}
\caption{Top: Star-formation rate versus stellar mass, colour-coded by infrared template type. The colour coding is the same as in Fig. 16. Black lines show e-folding times for stellar mass of 1, 10 and 100 Gyr. It is clear that most galaxies lie below the 10 Gyr line, indicating that they must have formed stars at a higher rate in the past. The red broken line is the ``main-sequence" locus of Elbaz et al. (2011). Bottom: Dust mass versus stellar mass.  The straight lines correspond, from the right, to  $M_{\rm dust}/M_* = 10^{-6}, 10^{-4}$ and $10^{-2}$.}
\label{plotsfrmstarmdust}
\end{figure}

We have followed the methodology of Rowan-Robinson et al (2008) to calculate stellar masses, star-formation rates (SFR) and dust masses for the sources in our catalogue.  For the sake of completeness, we briefly summarise our methodology here. We estimate the rest-frame 3.6\ $\micron$ luminosity $\nu L_{\nu}(3.6)$ which is then converted to stellar mass $M_*$ using the ratio $(M_*/M_{\odot})/(\nu L_{\nu}(3.6)/L_{\odot})$ derived from stellar synthesis models. To estimate star-formation rate, we use the conversion recipes of Rowan-Robinson et al. (1997) and Rowan-Robinson (2001), 
\begin{equation}
{\rm SFR}= 2.2 \epsilon^{-1} 10^{-10} (L_{60}/L_{\odot})
\end{equation}
where $\epsilon$ is the fraction of UV light absorbed by dust, taken as 2/3. 
To estimate the approximate dust mass for each galaxy, we use radiative transfer models for cirrus, M82 and A220 starburst components and the recipe
\begin{equation}
M_{\rm dust} /M_{\odot} = k L_{\rm IR} / L_{\odot}
\end{equation}
where $k=1.3\times10^{-3}$, $1.1\times10^{-4}$ and $4.4\times10^{-4}$ for cirrus, M82 and A220 respectively (A. Efstathiou, 2007, private communication).
The left panel in Fig.~\ref{plotsfrmstarmdust} shows a plot of SFR ($M_{\odot}/yr$) versus stellar mass ($M_{\odot}$).  Straight line loci correspond to the time-scales to make this mass of stars, forming stars at this rate.  While many galaxies are forming stars at a rate broadly consistent with that required to generate the observed stellar mass over $\sim$ 10 Gyr, most local galaxies lie to the right of the 10 Gyr line and must have formed stars at a higher rate at some time in the past.  The zone occupied by our complete and all-sky sample of local quiescent galaxies, lies to the right of the ``main-sequence" locus of Elbaz et al (2011) and of the 10 Gyr line, consistent with the fact that these galaxies are forming stars at a much lower rate than they did in the past (at $z = 1-3$).  Galaxies in Fig.~\ref{plotsfrmstarmdust} have been colour-coded by their dominant infrared template type as in Fig.~\ref{plot350550errir6ir11ht}. The quiescent, optically thin galaxies, i.e. those whose SEDs are dominated by cirrus, lie on the right edge of the distribution, while galaxies whose infrared SEDs are dominated by starbursts lie to the left, with higher specific star-formation rate.  The right panel in Fig.~\ref{plotsfrmstarmdust} shows dust mass versus stellar mass, with most galaxies having $M_{\rm dust}/M_*$ in the range $10^{-6}$ to $10^{-2}$. Galaxies dominated by cool cirrus templates tend to require higher specific dust masses.



\section{Conclusions and discussions}

\begin{figure*}
\includegraphics[height=2.95in,width=5.4in]{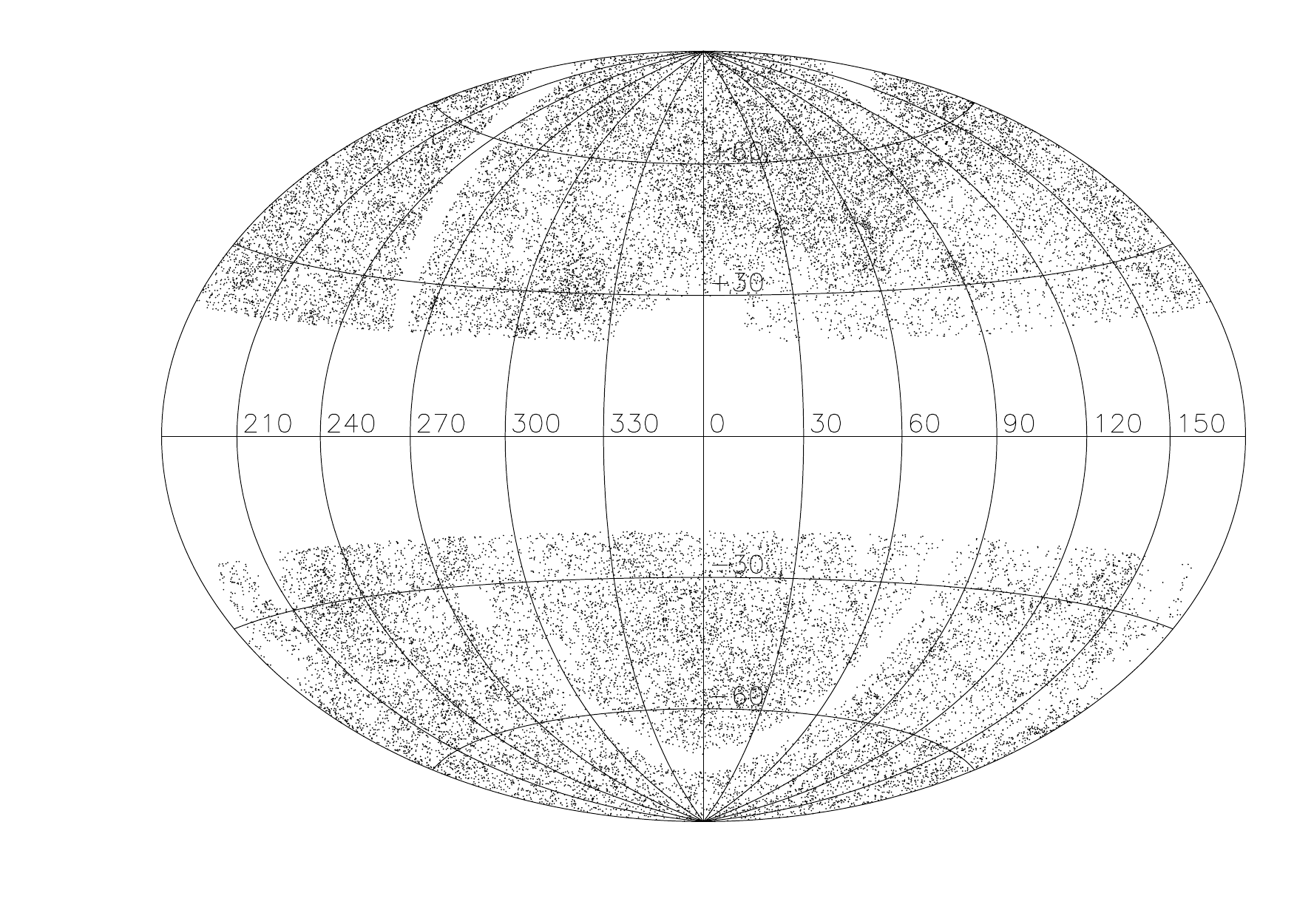}
\includegraphics[height=2.95in,width=5.4in]{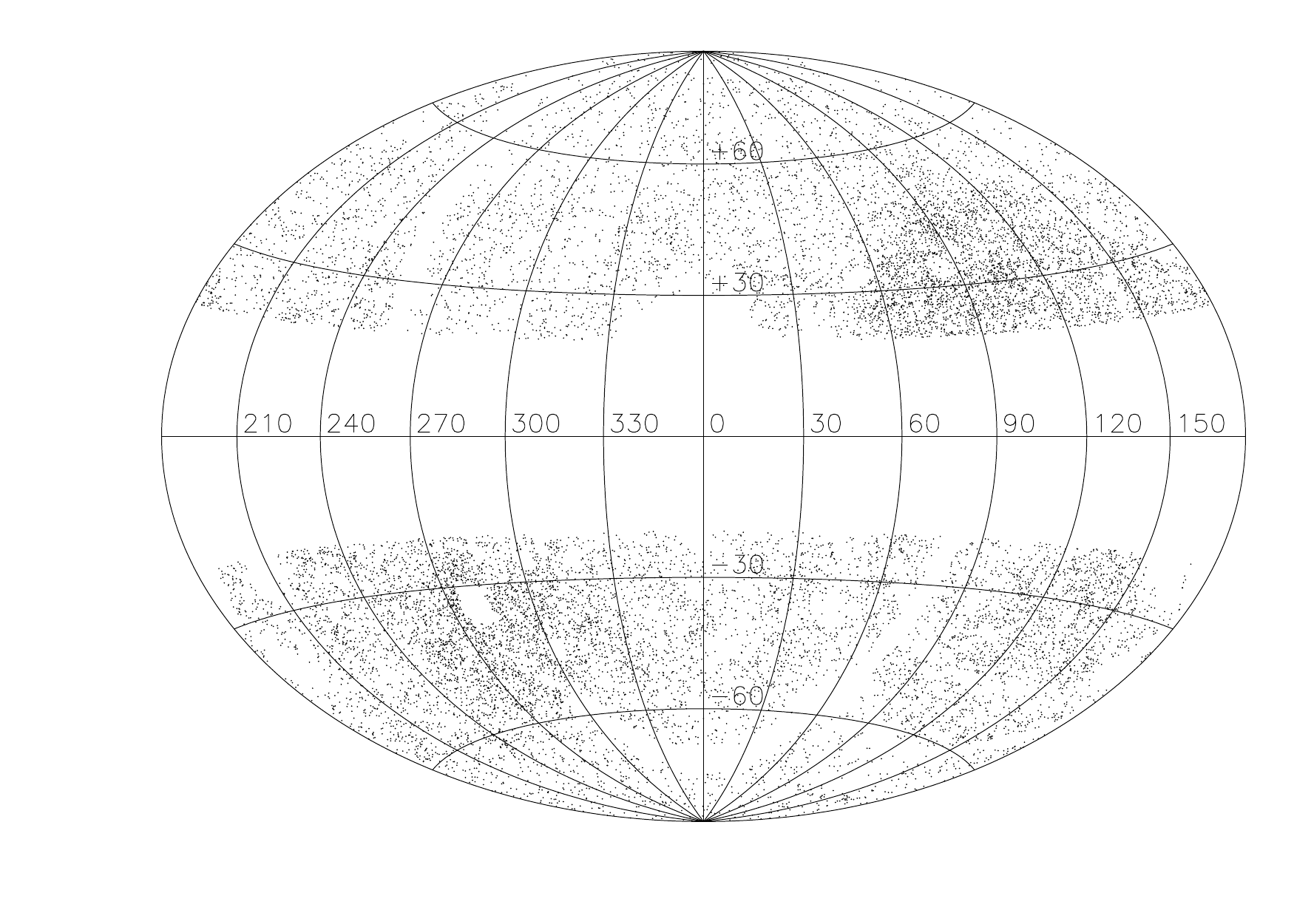}
\includegraphics[height=2.95in,width=5.4in]{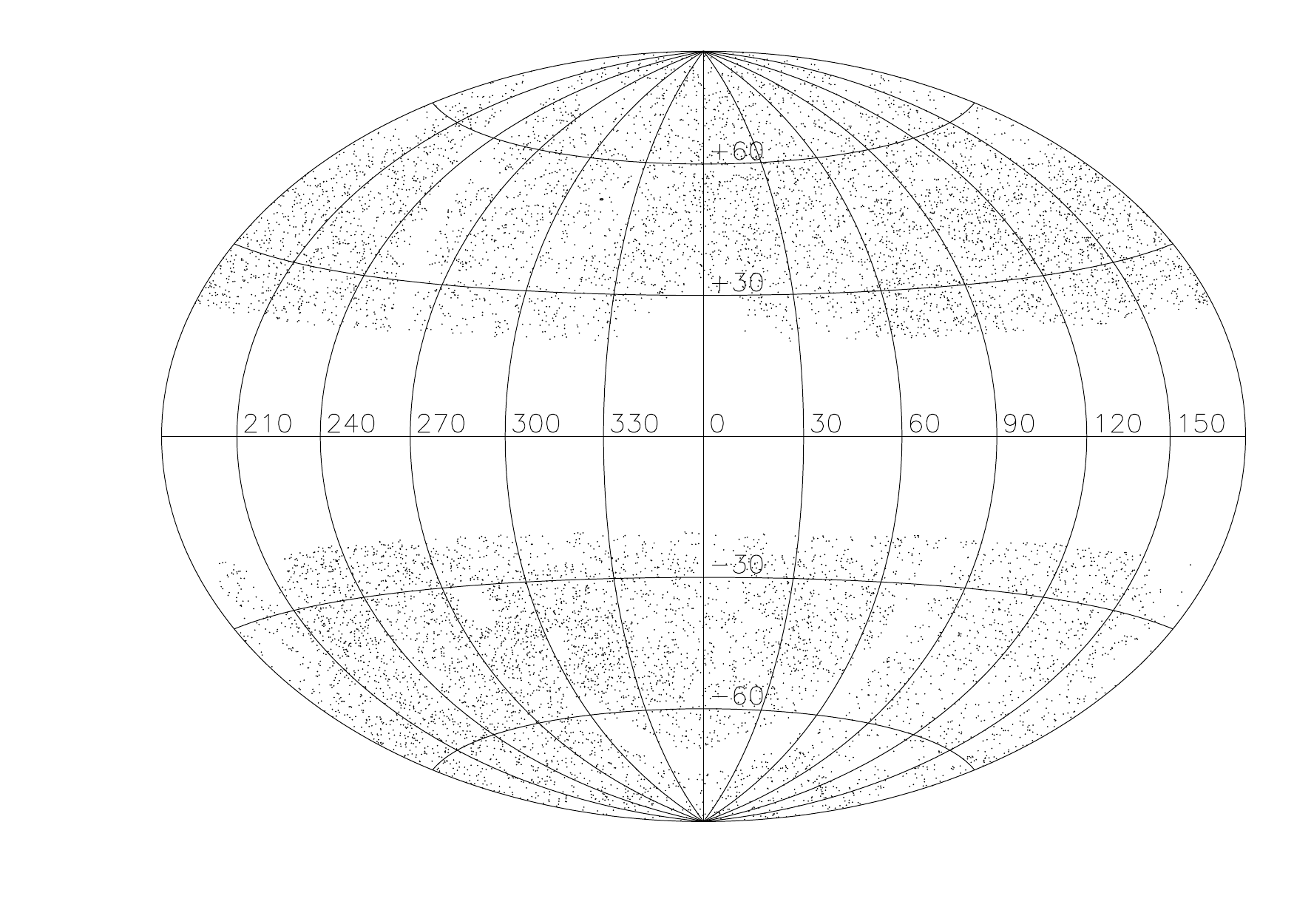}
\caption{The sky distribution (in Galactic coordinates) of all galaxies in the RIFSCz which have received spectroscopic redshifts (top), all galaxies with photometric redshift but without spectroscopic redshifts (middle), and all galaxies with neither spectroscopic or photometric redshifts (bottom).}
\label{skydistribution}
\end{figure*}

To summarise, the RIFSCz contains a total of 60,303 galaxies, $56\%$ of which have spectroscopic redshifts from NED, FSSz, PSCz, 6dF and the SDSS spectroscopic DR10 survey and $26\%$ of which have photometric redshifts from the template-fitting method. At a flux limit of S60 = 0.36 Jy which is the $90\%$ completeness limit of the FSC, more than $93\%$ of the galaxies in the RIFSCz have either spectroscopic or photometric redshifts.The RIFSCz covers about $61\%$ of the whole sky. In Fig.~\ref{skydistribution}, the sky distribution in Galactic coordinates is plotted for all galaxies in the RIFSCz with either spectroscopic or photometric redshift and galaxies without any redshift estimate. 

\begin{figure}
\includegraphics[height=3.in,width=3.45in]{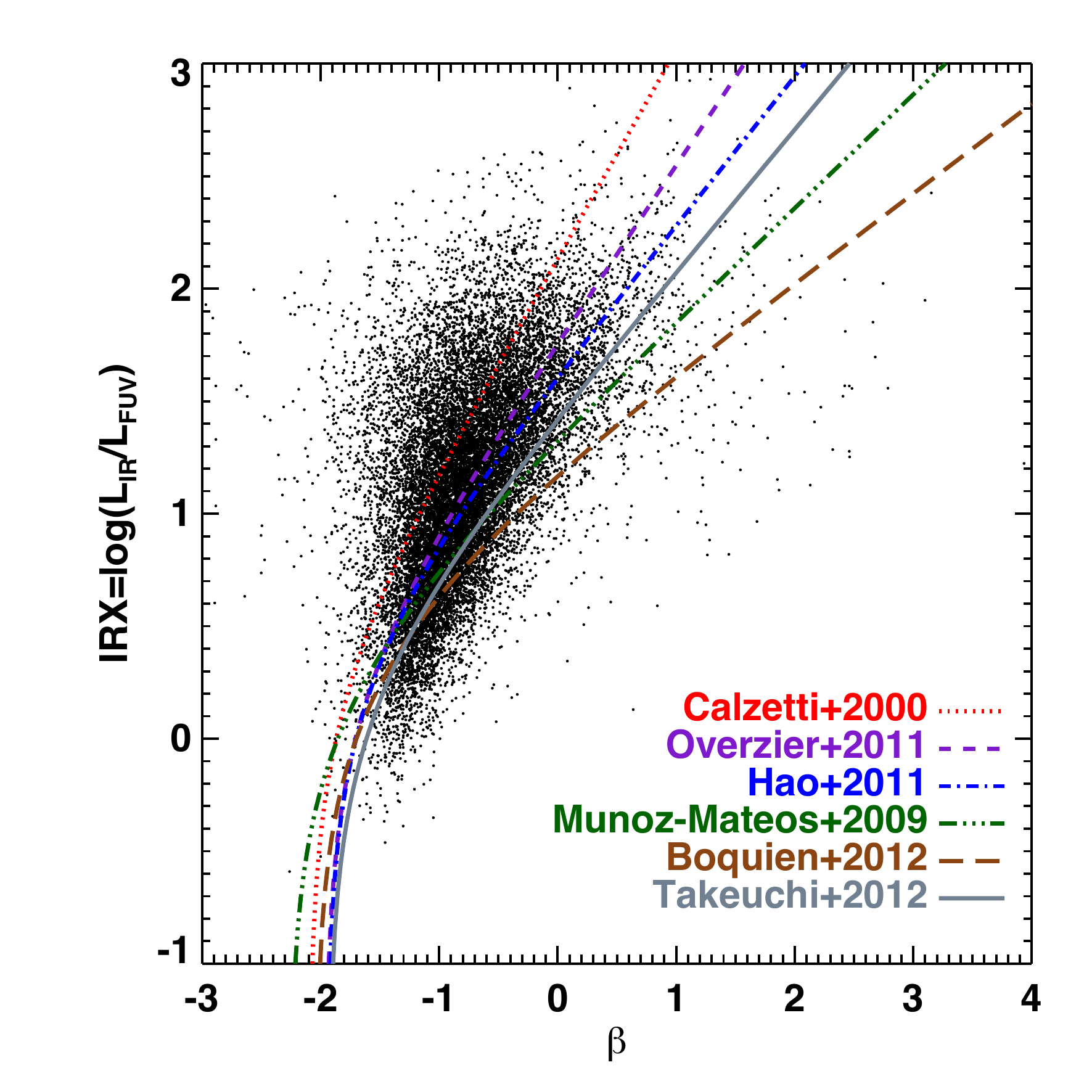}
\includegraphics[height=3.in,width=3.45in]{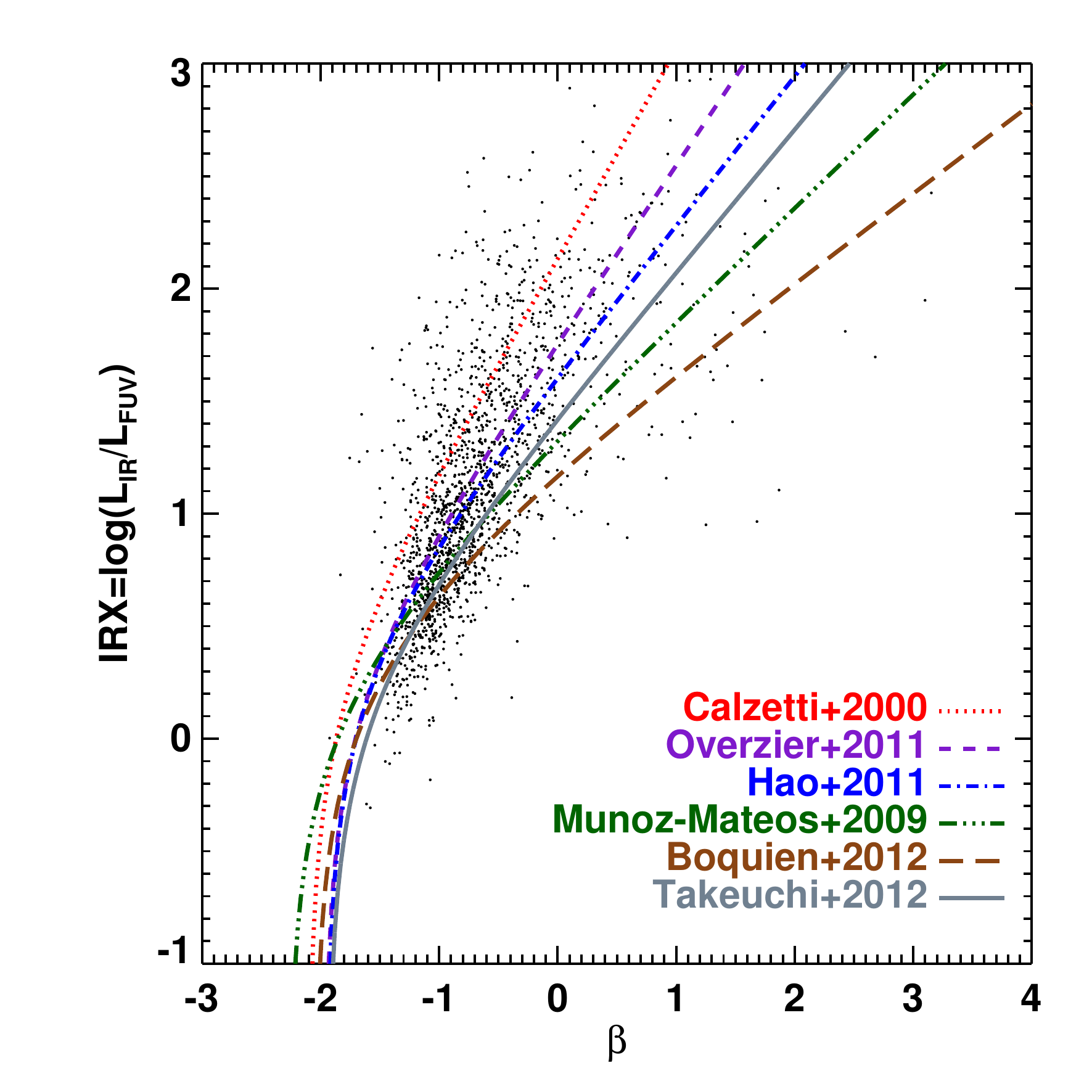}
\caption{The infrared to UV luminosity (the infrared excess: IRX) versus the slope of the UV spectrum ($\beta$) for galaxies in the RIFSCz catalogue (top: galaxies with SNR $\ge5$ UV detections and the reduced $\chi^2$ of the infrared SED fit $\le3$; bottom: a subset of the galaxies in the top panel which are detected at 350\ $\micron$.), compared with other relations published in the literature. The typical statistical error of the infrared luminosity ($L_{\rm IR}$) for galaxies in the top and bottom panel is about 0.3 dex and 0.1 dex respectively.}
\label{irx_beta}
\end{figure}

The RIFSCz catalogue represents the best all-sky far-infrared (60\ $\micron$) selected galaxy catalogue with multi-wavelength photometric information. As a result, several extragalactic science programmes can be carried out with the RIFSCz dataset. For example, we can examine the relation between the ratio of the infrared to UV luminosity  (the infrared excess: IRX) and the slope of the UV spectrum ($\beta$). This relation is critical to estimating the dust attenuation in galaxies and hence the true star-formation rate. In the top panel in Fig.~\ref{irx_beta}, we plot IRX versus $\beta$ for RIFSCz galaxies with GALEX UV detections at SNR $\ge5$ and the reduced $\chi^2$ of the infrared SED fit $\le3$. The typical statistical error of the infrared luminosity ($L_{\rm IR}$) for these galaxies is about 0.3 dex.  In the bottom panel in Fig.~\ref{irx_beta}, we plot IRX versus $\beta$ for a subset of the galaxies in the top panel which are detected at 350\ $\micron$ by Planck. The typical statistical error of $L_{\rm IR}$ for this subset of galaxies with 350\ $\micron$ is around 0.1 dex. Our $\beta$ values are derived using the recipe from Kong et al. (2004). A compilation of the IRX - $\beta$ relations published in the literature is also plotted. The Calzetti et al. (2000) and Overzier et al. (2011) relations are derived from UV-selected samples. The Boquien et al. (2012) relation is derived from a $K$-band selected sample. The Mu{\~n}oz-Mateos et al. (2009) relation is derived from the SINGS sample (Kennicutt et al. 2003) which consists of 75 local galaxies spanning a large range in morphology, luminosity, SFR and opacity.  The Hao et al. (2011) relation is derived from the SINGS sample combined with a sample of local galaxies covering the full range of optical spectral characteristics (Moustakas \& Kennicutt 2006). It is clear that there is a large dispersion among the different relations derived from samples with different selection criteria and our far-infrared selected sample show a general good agreement with these relations. In addition, our far-infrared selected sample also seems to cover regions with high IRX but low $\beta$ values which are not represented in other relations.  Detailed investigations are required to fully understand the causes of the difference between our IRX - $\beta$ relation based on far-infrared selection and the other relations. However, it is beyond the scope of this paper.


We provide a long version and a short version of the RIFSCz to the public, available from \url{http://astro.ic.ac.uk/public/mrr/fss/readme} and \url{http://www.astro.dur.ac.uk/~lwang83/RIFSCz.tar.gz}. The long version contains all the information (e.g. positional, photometric or spectroscopic) we have either assembled or derived in the process of constructing the RIFSCz. Then short version contains the most important information of the sources in the RIFSCz and should suffice for most users. The format of the long and short version of the RIFSCz catalogue is explained in detail in the Appendix.

We have presented the revised IRAS-FSC Redshift Catalogue (RIFSCz). It contains 60,303 galaxies selected at 60\ $\micron$ from the IRAS Faint Source Catalog, covering around $61\%$ of the whole sky. The process of retrieving spectroscopic redshifts, multi-wavelength source identification and photometric redshift estimation is described in detail.  Users should be aware of issues such as the intrinsic IRAS FSC completeness limit, the redshift completeness variations across the sky and the varying quality of photometric redshifts derived for different subsets of the catalogue.

\section*{ACKNOWLEDGEMENTS}

LW and PN acknowledge support from an ERC StG grant (DEGAS-259586). PN also acknowledges the support of the Royal Society through the award of a University Research Fellowship.

The NASA/IPAC EXTRAGALACTIC DATABASE (NED) is operated by the JET PROPULSION LABORATORY, CALTECH, under contract with the NATIONAL AERONAUTICS AND SPACE ADMINISTRATION.

Funding for the Sloan Digital Sky Survey (SDSS) and SDSS-II has been provided by the Alfred P. Sloan Foundation, the Participating Institutions, the National Science Foundation, the U.S. Department of Energy, the National Aeronautics and Space Administration, the Japanese Monbukagakusho, and the Max Planck Society, and the Higher Education Funding Council for England. The SDSS Web site is \url{http://www.sdss.org/}.

The SDSS is managed by the Astrophysical Research Consortium (ARC) for the Participating Institutions. The Participating Institutions are the American Museum of Natural History, Astrophysical Institute Potsdam, University of Basel, University of Cambridge, Case Western Reserve University, The University of Chicago, Drexel University, Fermilab, the Institute for Advanced Study, the Japan Participation Group, The Johns Hopkins University, the Joint Institute for Nuclear Astrophysics, the Kavli Institute for Particle Astrophysics and Cosmology, the Korean Scientist Group, the Chinese Academy of Sciences (LAMOST), Los Alamos National Laboratory, the Max-Planck-Institute for Astronomy (MPIA), the Max-Planck-Institute for Astrophysics (MPA), New Mexico State University, Ohio State University, University of Pittsburgh, University of Portsmouth, Princeton University, the United States Naval Observatory, and the University of Washington.

Funding for SDSS-III has been provided by the Alfred P. Sloan Foundation, the Participating Institutions, the National Science Foundation, and the U.S. Department of Energy Office of Science. The SDSS-III web site is http://www.sdss3.org/.

SDSS-III is managed by the Astrophysical Research Consortium for the Participating Institutions of the SDSS-III Collaboration including the University of Arizona, the Brazilian Participation Group, Brookhaven National Laboratory, Carnegie Mellon University, University of Florida, the French Participation Group, the German Participation Group, Harvard University, the Instituto de Astrofisica de Canarias, the Michigan State/Notre Dame/JINA Participation Group, Johns Hopkins University, Lawrence Berkeley National Laboratory, Max Planck Institute for Astrophysics, Max Planck Institute for Extraterrestrial Physics, New Mexico State University, New York University, Ohio State University, Pennsylvania State University, University of Portsmouth, Princeton University, the Spanish Participation Group, University of Tokyo, University of Utah, Vanderbilt University, University of Virginia, University of Washington, and Yale University.

This publication makes use of data products from the Two Micron All Sky Survey, which is a joint project of the University of Massachusetts and the Infrared Processing and Analysis Center/California Institute of Technology, funded by the National Aeronautics and Space Administration and the National Science Foundation.

This publication makes use of data products from the Wide-field Infrared Survey Explorer, which is a joint project of the University of California, Los Angeles, and the Jet Propulsion Laboratory/California Institute of Technology, funded by the National Aeronautics and Space Administration.

\appendix

\section{The long version of the RIFSCz catalogue}

Notes on the catalogue columns for the long version:

Col 1 - 7. IRAS FSC source name, GALEX source ID, SDSS object ID, 2MASS XSC source designation, WISE source designation, Planck source name and NED source name.

Col 8 - 10. SDSS type, NED type, stellar flag (1= galaxy/extended,  -1 = stellar/QSO, and 0 = unknown).

Col 11 - 22. IRAS FSC positions, SDSS positions, 2MASS XSC positions, WISE positions, Planck positions and NED positions.

Col 23 - 25. Recommended position and positional flag (1=SDSS, 2=2MASS, 3=WISE, 4=NED, 5=FSC and prioritised in the same order).

Col 26 - 28. SDSS spectroscopic redshift, 2MRS redshift, and NED redshift.

Col 29 - 30. Recommended spectroscopic redshift and flag (1$=$SDSS, 2$=$PSCz, 3$=$FSSz, 4$=$6dF, 5$=$NED and 6$=$2MRS). The priority order is NED$>$SDSS$>$2MRS$>$PSCz$>$FSSz$>$6dF. 

Col 31. Template-fitting photometric redshift for sources with optical or near-infrared photometry

Col 32. Recommended redshift which is equal to spectroscopic redshift if it is available or photometric redshift if it has optical or near-infrared photometry but without spectroscopic redshift.

Col 33 - 37. The number of optical and near-infrared bands used in the photometric redshift fit, optical galaxy template type, extinction AV from the optical galaxy template fit, reduced $\chi^2$ of the best redshift fit, and the absolute B magnitude.


Col 38 - 41. GALEX far-UV flux and error,  GALEX  near-UV flux and error.

Col 42 - 51. SDSS  magnitudes and errors at u, g, r, i, z

Col 52 - 57. 2MASS PSC  magnitudes and errors at J, H, K$_s$.

Col 58 - 63. 2MASS XSC  magnitudes and errors at J, H, K$_s$.

Co 64 - 71. WISE magnitudes and errors at 3.4, 4.6, 12 and 22\ $\micron$, 

Col 72 - 79. IRAS fluxes and quality flags at 12, 25, 60 and 100\ $\micron$, 

Col 80 - 87. Planck fluxes and errors at 350, 550, 850 or 1382\ $\micron$. 

Col 88 - 89. Integrated 1.4 GHz flux density and error.

Col 90. Aperture correction.

Col 91 - 102. Aperture  corrected ugriz, 2MASS J, H, and K$_s$, and WISE 3.4, 4.6, 12 and 22\ $\micron$ fluxes. 

Col 103 - 106. Redshift flag (1=spectroscopic redshift, 2=photometric redshift, 3=no redshift), Infrared template type (1=cirrus, 2=M82, 3=A220, 4=AGN dust torus, 5=young starburst, 6=cool cirrus), reduced $\chi^2$ for the IR template fit, number of photometric bands used in the IR template fit

Col 107 - 112. Fraction of contribution at 60\ $\micron$ of each of the six IR templates. 

Col 113 - 118. Fraction of IR luminosity in each of the six IR templates 

Col 119. Total IR luminosity. 

Col 120 - 133. Predicted infrared and sub-mm fluxes at 12, 25, 60, 90, 100, 110, 140, 160, 250, 350, 550, 850, 1250 and 1380\ $\micron$ based on the best-fit infrared template.

Col 134 - 136. Stellar mass, SFR,  dust mass.

Col 137 - 221 array of reduced $\chi^2$, minimised over all templates, in bins of 0.01 in $\log_{10}(1+z)$, from 0 to 0.85.










\section{The short version of the RIFSCz catalogue}

Notes on the catalogue columns for the short version:

Col 1. IRAS FSC source name.

Col 2. Stellar flag (1= galaxy/extended,  -1 = stellar/QSO, and 0 = unknown).

Col 3 - 5. Recommended position and positional flag (1=SDSS, 2=2MASS, 3=WISE, 4=NED, 5=FSC and prioritised in the same order).

Col 6 - 7. Recommended spectroscopic redshift and flag (1$=$SDSS, 2$=$PSCz, 3$=$FSSz, 4$=$6dF, 5$=$NED and 6$=$2MRS). The priority order is NED$>$SDSS$>$2MRS$>$PSCz$>$FSSz$>$6dF. 

Col 8. Template-fitting photometric redshift for sources with optical or near-infrared photometry

Col 9. Recommended redshift which is equal to spectroscopic redshift if it is available or photometric redshift if it has optical or near-infrared photometry but without spectroscopic redshift.

Col 10 - 14. The number of optical and near-infrared bands used in the photometric redshift fit, optical galaxy template type, extinction AV from the optical galaxy template fit, reduced $\chi^2$ of the best redshift fit, and the absolute B magnitude.

Col 15 - 18. GALEX far-UV flux and error,  GALEX  near-UV flux and error.

Col 19 - 28. SDSS  magnitudes and errors at u, g, r, i, z

Col 29 - 34. 2MASS PSC  magnitudes and errors at J, H, K$_s$.

Col 35 - 40. 2MASS XSC  magnitudes and errors at J, H, K$_s$.

Co 41 - 48. WISE magnitudes and errors at 3.4, 4.6, 12 and 22\ $\micron$, 

Col 49 - 56. IRAS fluxes and quality flags at 12, 25, 60 and 100\ $\micron$, 

Col 57 - 64. Planck fluxes and errors at 350, 550, 850 or 1382\ $\micron$. 

Col 65 - 66. Integrated 1.4 GHz flux density and error.

Col 67 - 78. Aperture  corrected ugriz, 2MASS J, H, and K$_s$, and WISE 3.4, 4.6, 12 and 22\ $\micron$ fluxes. 

Col 79 - 82. Redshift flag (1=spectroscopic redshift, 2=photometric redshift, 3=no redshift), Infrared template type (1=cirrus, 2=M82, 3=A220, 4=AGN dust torus, 5=young starburst, 6=cool cirrus), reduced $\chi^2$ for the IR template fit, number of photometric bands used in the IR template fit

Col 83 - 88. Fraction of contribution at 60\ $\micron$ of each of the six IR templates. 

Col 89 - 94. Fraction of IR luminosity in each of the six IR templates 

Col 95. Total IR luminosity. 

Col 96 - 109. Predicted infrared and sub-mm fluxes at 12, 25, 60, 90, 100, 110, 140, 160, 250, 350, 550, 850, 1250 and 1380\ $\micron$ based on the best-fit infrared template.

Col 110 - 112. Stellar mass, SFR,  dust mass.


\begin{thebibliography}{99}


\bibitem[Ahn et al.(2013)]{ahn13} 
Ahn, C.~P., Alexandroff, R., Allende Prieto, C., et al.\ 2013, arXiv:1307.7735 

\bibitem[Boquien et al.(2012)]{boq12} 
Boquien, M., Buat, V., Boselli, A., et al.\ 2012, \aap, 539, A145 


\bibitem[Brusa et al.(2007)]{bru07} 
Brusa, M., Zamorani, G., Comastri, A., et al.\ 2007, \apjs, 172, 353 


\bibitem[Budav{\'a}ri et al.(2009)]{bud09} 
Budav{\'a}ri, T., Heinis, S., Szalay, A.~S., et al.\ 2009, \apj, 694, 1281 

\bibitem[Calzetti et al.(2000)]{cal00} 
Calzetti, D., Armus, L., Bohlin, R.~C., et al.\ 2000, \apj, 533, 682 


\bibitem[Chapin et al.(2011)]{cha11} 
Chapin, E.~L., Chapman, S.~C., Coppin, K.~E., et al.\ 2011, \mnras, 411, 505 
 
 \bibitem[Efstathiou et al.(2000)]{efs00} 
 Efstathiou, A., Rowan-Robinson, M., \& Siebenmorgen, R.\ 2000, \mnras, 313, 734

\bibitem[Efstathiou \& Rowan-Robinson(2003)]{efs03} 
Efstathiou, A., \& Rowan-Robinson, M.\ 2003, \mnras, 343, 322 

 

\bibitem[Elbaz et al.(2011)]{elb11} 
Elbaz, D., Dickinson, M., Hwang, H.~S., et al.\ 2011, \aap, 533, A119 

\bibitem[Hao et al.(2011)]{hao11} 
Hao, C.-N., Kennicutt, R.~C., Johnson, B.~D., et al.\ 2011, \apj, 741, 124 

 \bibitem[Heinis et al.(2009)]{hei09} 
Heinis, S., Budav{\'a}ri, T., \& Szalay, A.~S.\ 2009, \apj, 705, 739 

\bibitem[Huchra et al.(2012)]{huc12} 
Huchra, J.~P., Macri, L.~M., Masters, K.~L., et al.\ 2012, \apjs, 199, 26 
 

\bibitem[Jarrett et al.(2000)]{jar00} 
Jarrett, T.~H., Chester, T., Cutri, R., et al.\ 2000, \aj, 119, 2498 


 \bibitem[Jones et al.(2004)]{jon04} 
 Jones, D.~H., Saunders, W., Colless, M., et al.\ 2004, \mnras, 355, 747 


 \bibitem[Jones et al.(2005)]{jon05} 
 Jones, D.~H., Saunders, W., Read, M., \& Colless, M.\ 2005, PASA, 22, 277 


\bibitem[Kennicutt et al.(2003)]{ken03} 
Kennicutt, R.~C., Jr., Armus, L., Bendo, G., et al.\ 2003, \pasp, 115, 928 

\bibitem[Kong et al.(2004)]{kon04} 
Kong, X., Charlot, S., Brinchmann, J., \& Fall, S.~M.\ 2004, \mnras, 349, 769 

 \bibitem[Moshir et al.(1992)]{mos92} 
 Moshir, M., Kopman, G., \& Conrow, T.~A.~O.\ 1992, Pasadena: Infrared Processing and Analysis Center, California Institute of Technology, 1992, edited by Moshir, M.; Kopman, G.; Conrow, T.~a.o.,  


\bibitem[Moustakas \& Kennicutt(2006)]{mou06} 
Moustakas, J., \& Kennicutt, R.~C., Jr.\ 2006, \apjs, 164, 81 



\bibitem[Mu{\~n}oz-Mateos et al.(2009)]{mun09} 
Mu{\~n}oz-Mateos, J.~C., Gil de Paz, A., Boissier, S., et al.\ 2009, \apj, 701, 1965 



\bibitem[Overzier et al.(2011)]{ove11} 
Overzier, R.~A., Heckman, T.~M., Wang, J., et al.\ 2011, \apjl, 726, L7 


 \bibitem[Planck Collaboration et al.(2011)]{pla11} 
 Planck Collaboration, Ade, P.~A.~R., Aghanim, N., et al.\ 2011, \aap, 536, A16 

 \bibitem[Planck Collaboration et al.(2013)]{pla13a} 
 Planck Collaboration, Ade, P.~A.~R., Aghanim, N., et al.\ 2013, arXiv:1303.5062 


 \bibitem[Planck Collaboration et al.(2013)]{pla13b} 
 Planck Collaboration, Ade, P.~A.~R., Aghanim, N., et al.\ 2013, arXiv:1303.5088 


\bibitem[Rice et al.(1988)]{Ric88} 
Rice, W., Lonsdale, C.~J., Soifer, B.~T., et al.\ 1988, \apjs, 68, 91 


\bibitem[Rodighiero et al.(2011)]{rod11} 
Rodighiero, G., Daddi, E., Baronchelli, I., et al.\ 2011, \apjl, 739, L40 


\bibitem[Rowan-Robinson et al.(2005)]{row05} 
Rowan-Robinson, M., Babbedge, T., Surace, J., et al.\ 2005, \aj, 129, 1183 


\bibitem[Rowan-Robinson et al.(2008)]{row08} 
Rowan-Robinson M. et al, 2008, MNRAS 386, 687

\bibitem[Rowan-Robinson \& Efstathiou(2009)]{row09} 
Rowan-Robinson M. and Efstathiou A., 2009, MNRAS 399, 615

\bibitem[Rowan-Robinson \& Wang(2010)]{row10a} 
Rowan-Robinson, M., \& Wang, L.\ 2010, \mnras, 406, 720 

\bibitem[Rowan-Robinson et al.(2010)]{row10b} 
Rowan-Robinson, M., Roseboom, I.~G., Vaccari, M., et al.\ 2010, \mnras, 409, 2 

\bibitem[Rowan-Robinson et al.(2013)]{row13}
Rowan-Robinson M. et al, 2013, MNRAS 428, 1958

\bibitem[Salpeter(1955)]{sal55} 
Salpeter, E.~E.\ 1955, \apj, 121, 161 


\bibitem[Saunders et al.(2000)]{sau00} 
Saunders, W., Sutherland, W.~J., Maddox, S.~J., et al.\ 2000, \mnras, 317, 55 




\bibitem[Skrutskie et al.(2006)]{skr06} 
Skrutskie, M.~F., Cutri, R.~M., Stiening, R., et al.\ 2006, \aj, 131, 1163 

\bibitem[Spoon et al.(2007)]{spo07} 
Spoon, H.~W.~W., Marshall, J.~A., Houck, J.~R., et al.\ 2007, \apjl, 654, L49 


\bibitem[Sutherland \& Saunders(1992)]{sut92} 
Sutherland, W., \& Saunders, W.\ 1992, \mnras, 259, 413 



\bibitem[Takeuchi et al.(2012)]{tak12} 
Takeuchi, T.~T., Yuan, F.-T., Ikeyama, A., Murata, K.~L., \& Inoue, A.~K.\ 2012, \apj, 755, 144 

\bibitem[Wang \& Rowan-Robinson(2009)]{wan09} 
Wang, L., \& Rowan-Robinson, M.\ 2009, \mnras, 398, 109

\bibitem[Wang \& Rowan-Robinson(2010)]{wan10} 
Wang, L., \& Rowan-Robinson, M.\ 2010, \mnras, 401, 35 

\bibitem[Wang et al.(2013)]{wan13} 
Wang, L., et al. 2013, arxiv:1312.0552


\bibitem[Wright et al.(2010)]{wri10} 
Wright, E.~L., Eisenhardt, P.~R.~M., Mainzer, A.~K., et al.\ 2010, \aj, 140, 1868 

\bibitem[Yan et al.(2013)]{yan13} 
Yan, L., Donoso, E., Tsai, C.-W., et al.\ 2013, \aj, 145, 55 


\bibitem[York et al.(2000)]{yor00} 
York, D.~G., Adelman, J., Anderson, J.~E., Jr., et al.\ 2000, \aj, 120, 1579 




\end{thebibliography}
\end{document}